\documentclass[iop]{emulateapj}
\usepackage{amsmath,graphicx,times,bm,url,xspace,color,enumitem}
\pdfoutput=1
\defcitealias{2013ApJ...769...76B}{BS13}
\newcommand{\BS}{\citetalias{2013ApJ...769...76B}\xspace}

\usepackage{ulem}

\renewcommand\emph[1]{\textit{#1}}

\newcommand\cm{\,\rm cm}
\newcommand\m{\,\rm m}
\newcommand\s{\,\rm s}

\newcommand\g{\,\rm g}
\newcommand\erg{\,\rm erg}
\newcommand\K{\,\rm K}
\newcommand\yr{\,\rm yr}
\newcommand\Myr{\,\rm Myr}

\newcommand\km{\,\rm km}
\newcommand\kms{\,\rm km\,s^{-1}}

\newcommand\au{\,\rm au}
\newcommand\mG{\,\rm mG}

\newcommand\Msun{\,M_\odot}

\newcommand\yes{$\circ$}
\newcommand\no{$-$}

\newcommand\tms{\!\times\!}
\newcommand\cdt{\!\cdot\!}

\newcommand\rr{\mathbf r}

\newcommand\bb{\hat{{\mathbf b}}}

\newcommand\V{\mathbf v}
\newcommand\B{\mathbf B}

\newcommand\etao{\eta_{\rm O}}
\newcommand\etad{\eta_{\rm AD}}
\newcommand\EO{\mathbf E_{\rm O}}
\newcommand\EAD{\mathbf E_{\rm AD}}

\newcommand\Rc{R}
\newcommand\cS{c_{\rm s}}

\newcommand\Rm{\mathrm{Rm}}

\newcommand{\simgt}%
           {\,\hbox{\lower0.35ex\hbox{$\sim$}\llap{\raise0.35ex\hbox{$>$}}}\,}
\newcommand{\simlt}%
           {\,\hbox{\lower0.35ex\hbox{$\sim$}\llap{\raise0.35ex\hbox{$<$}}}\,}

\newcommand\NIII{\textsc{nirvana-iii}\xspace}

\begin{document}

\title{Global simulations of protoplanetary disks with Ohmic
  resistivity and ambipolar~diffusion}

\author{
  Oliver~Gressel$^{1}$, Neal~J.~Turner$^{2}$, 
  Richard~P.~Nelson$^{3}$ and Colin~P.~McNally$^1$
}

\affil{
  $^1$Niels Bohr International Academy, The Niels Bohr Institute, 
      Blegdamsvej 17, DK-2100, Copenhagen \O, Denmark\\
  $^2$Jet Propulsion Laboratory, California Institute of
      Technology, Pasadena, CA 91109, USA\\
  $^3$Astronomy Unit, Queen Mary University of London,
      Mile End Road, London E1 4NS, UK
}

\email{
  oliver.gressel@nbi.dk (OG);
  neal.turner@jpl.nasa.gov (NJT);
  r.p.nelson@qmul.ac.uk (RPN);
  cmcnally@nbi.dk (CPMN)
}

% ------------------------------------------------------------------------------

\begin{abstract}
%context
Protoplanetary disks are believed to
accrete onto their central T~Tauri star because of magnetic
stresses. Recently published shearing box simulations indicate that
Ohmic resistivity, ambipolar diffusion and the Hall effect all play
important roles in disk evolution. In the presence of a vertical
magnetic field, the disk remains laminar between 1--$5\au$, and a
magnetocentrifugal disk wind forms that provides an important
mechanism for removing angular momentum.
%aims
Questions remain, however, about the establishment of a true physical
wind solution in the shearing box simulations because of the
symmetries inherent in the local approximation.
%methods
We present global MHD simulations of protoplanetary disks that include
Ohmic resistivity and ambipolar diffusion, where the time-dependent
gas-phase electron and ion fractions are computed under FUV and X-ray
ionization with a simplified recombination chemistry.
%results
Our results show that the disk remains laminar, and that a physical
wind solution arises naturally in global disk models. The wind is
sufficiently efficient to explain the observed accretion rates.
Furthermore, the ionization fraction at intermediate disk heights is
large enough for magneto-rotational channel modes to grow and
subsequently develop into belts of horizontal field. Depending on the
ionization fraction, these can remain quasi-global, or break-up into
discrete islands of coherent field polarity.
%conclusions
The disk models we present here show a dramatic departure from our
earlier models including Ohmic resistivity only. It will be important
to examine how the Hall effect modifies the evolution, and to explore
the influence this has on the observational appearance of such
systems, and on planet formation and migration.
\end{abstract}


% ------------------------------------------------------------------------------

\section{Introduction}
\label{sec:intro}

Understanding the complex dynamical evolution of protoplanetary disks
(PPDs) is of key interest both for building a comprehensive theory of
planet formation, and as a means of explaining the observationally
inferred properties of these objects \citep[see][for a
recent review]{2014arXiv1401.7306T}. For example, PPDs are known
to accrete gas onto their host stars at a typical rate of $10^{-8 \pm
  1} \Msun \yr^{-1}$ \citep{1998ApJ...492..323G, 1998ApJ...495..385H}
and have life times in the range $\sim 3$--$10 \Myr$
\citep{2006ApJ...638..897S}.  More recent observations have indicated
the presence of turbulence in the disks surrounding the young stars
TW~Hydrae and HD~163296, based on analysis of their molecular line
emission \citep{2011ApJ...727...85H}.  Further evidence for turbulent
broadening comes, for instance, from the CO rovibrational band head
shape measured by \citet{2009ApJ...691..738N} in V1331 Cygni.  During
the T~Tauri (class II) phase, self-gravity is expected to provide a
negligible contribution to angular momentum transport (due to the low
disk-to-star mass ratio), and disk accretion is instead believed to be
driven by magnetic effects. Two possible sources of angular momentum
transport are through a magnetocentrifugally driven wind
\citep{1982MNRAS.199..883B, 1983ApJ...274..677P}, or through the
operation of the magnetorotational instability
\citep[MRI,][]{1991ApJ...376..214B}, whose nonlinear outcome in the
ideal-MHD limit is magnetohydrodynamic turbulence
\citep*{1991ApJ...376..223H,1995ApJ...440..742H}.

Except for the innermost regions of PPDs, where the temperature $T
\simgt 1000\K$, and the ionization of alkali metals such as sodium and
potassium provides strong coupling between the gas and magnetic field
\citep{1988PThPS..96..151U}, non-ideal MHD effects such as Ohmic
resistivity, ambipolar diffusion (AD) and Hall drift are expected to
be important due to the low ionization levels of the gas
\citep*[e.g.][]{1994ApJ...421..163B,1999MNRAS.307..849W,%
  2000ApJ...543..486S,2001ApJ...552..235B}.  External sources of
ionization such as stellar X-rays, FUV photons and galactic cosmic
rays are expected to ionize the disk surface layers, providing good
coupling there -- even though recent results suggest that cosmic rays
may be highly attenuated by the stellar wind
\citep{2013ApJ...772....5C}. Nevertheless, deep within the disk the
evolution should be controlled by the non-ideal effects. The
recognition of this led \cite{1996ApJ...457..355G} to propose
what has now become the
traditional dead-zone model in which the disk surface layers accrete
by sustaining MRI turbulence, with the shielded interior maintaining
an inert and magnetically decoupled ``dead zone'', where the MRI is
quenched by the action of Ohmic resistivity only. The potential
importance of the other non-ideal effects has long been recognized
\citep[e.g.][]{2002ApJ...577..534S, 2003MNRAS.345..992S}, but it is
only recently that nonlinear shearing box simulations including
ambipolar diffusion and the Hall term have been performed in relevant
parameter regimes, leading to a modified picture of how disks accrete
that deviates significantly from the traditional dead zone one.

Generally speaking, it is expected that in disk regions between
1--$5\au$, Ohmic resistivity will be dominant near the midplane, the
Hall effect will be dominant at intermediate disk heights and
ambipolar diffusion will be most important in low density regions
higher up in the disk \citep{2008MNRAS.388.1223S}.  The highest
altitude surface layers are expected to be ionized by stellar FUV
photons, and as such will evolve in the ideal MHD limit
\citep{2011ApJ...735....8P}.  Shearing box simulations presented by
\citet{2013ApJ...769...76B} that include Ohmic resistivity and
ambipolar diffusion for models computed at $1\au$ demonstrate that AD
quenches MRI turbulence. They conclude that disks will be completely
laminar between 1--$5\au$ with angular momentum transport occurring
because a magnetocentrifugal wind is launched from near the disk
surface.  For reasonable assumptions about the magnetic field strength
and geometry, accretion rates in agreement with the observations are
obtained. The inclusion of the Hall term in the presence of a vertical
magnetic field introduces dichotomous behavior, arising from the fact
that the coupled dynamics depends on the sign of ${\bf \Omega} \cdot
{\bf B}$, where ${\bf \Omega}$ and ${\bf B}$ signify the disk angular
momentum vector and the ambient magnetic field direction,
respectively. Shearing box simulations presented by
\cite{2014A&A...566A..56L} and \cite{2014ApJ...791..137B} show that
when ${\bf \Omega} \cdot {\bf B} > 0$, the Hall effect leads to the
amplification of the horizontal magnetic fields within approximately
two scale heights of the disk midplane, and the generation of a large
scale Maxwell stress through magnetic braking. In addition, the
magnetocentrifugally driven wind is also seen to strengthen. When
${\bf \Omega} \cdot {\bf B} < 0$, however, magnetic stresses and winds
are seen to weaken relative to the opposite case, and relative to the
evolution observed when Hall effects are neglected.  Quite how this
\emph{Hall dichotomy} maps onto observations of PPDs remains an open
question.

The internal dynamics of protoplanetary disks in the region 1--$5\au$
are clearly of great significance for planet formation. The growth and
settling of grains depends on the level of turbulence, with a laminar
disk or one displaying weak turbulence providing the most favorable
conditions -- although it should be noted that some turbulent mixing
is required to maintain a population of small grains in the surface
regions of PPDs so that their spectral energy distribution can be
reproduced by radiative transfer models \citep{2005A&A...434..971D}.
The collisional growth of planetesimals has been shown to be affected
strongly by the level of turbulence
\citep{2011MNRAS.415.3291G,2012MNRAS.422.1140G}, and the migration of
low mass planets is also highly sensitive to the presence or otherwise
of turbulence, with significant stress levels being required to
unsaturate corotation torques \citep{2011MNRAS.410..293P,
  2011A&A...533A..84B}.

This paper is the latest in a series in which we are exploring the
influence of magnetic fields and non ideal MHD effects on the
formation of planets, with the eventual goal of producing
comprehensive models of PPDs that will be used to study planet
building and evolution.  In earlier work
\citep{2010MNRAS.409..639N,2011MNRAS.415.3291G,2012MNRAS.422.1140G} we
have used a combination of global and shearing box simulations to
examine the dynamical evolution of dust grains, boulders and
planetesimals in turbulent disks, with and without dead zones. More
recently, we have studied the influence of magnetic fields on gap
formation and gas accretion onto a forming giant planet using global
simulations that also included Ohmic resistivity
\citep{2013ApJ...779...59G}.  In this paper, we include both Ohmic
resistivity and ambipolar diffusion in global disk simulations, and
follow the dynamical evolution of the resulting disk models as a
precursor to examining how ambipolar diffusion affects gas accretion
onto a giant planet. Of particular interest is the nature of the
accretion flow in the disk, the nature of any magnetocentrifugal wind
that is launched, and how these vary as a function of small changes in
model parameters.  These are the first quasi-global simulations to be
conducted of PPDs that are threaded by vertical magnetic fields and
which include this combination of non ideal MHD effects, and as such
are most comparable with the shearing box simulations presented by
\citet{2013ApJ...769...76B}, hereafter \BS. A fundamental question
raised by the shearing box simulations is whether or not a physical
wind solution with the correct field geometry emerges spontaneously in
global disk simulations in a way that is not generally observed in
shearing boxes because of their radial symmetry. Probably the most
important result contained in this paper is that physical wind
solutions do indeed arise spontaneously in our global simulations,
demonstrating the robustness of many of the features obtained in the
shearing box simulations of \BS.

This paper is organized as follows: In Section~\ref{sec:meth} we
describe the numerical method, the disk model, as well as the
ionization chemistry. Our results are presented in
Section~\ref{sec:res}, where we mainly focus on the emerging wind
solutions and the dynamical stability and evolution of forming current
layers. In Section~\ref{sec:corona}, we will moreover report an
instability that is driven by the sharp transition into the FUV
dominated layer, and in Section~\ref{sec:3d} we assess whether
secondary instabilities can drive non-axisymmetric evolution. We
summarize our results in Section~\ref{sec:summary} and discuss
implications for planet formation theory in
Section~\ref{sec:concl}.

% ------------------------------------------------------------------------------

\section{Methods}
\label{sec:meth}

We present magnetohydrodynamic (MHD) simulations of protoplanetary
accretion disks employing 2D axisymmetric and 3D uniformly-spaced
spherical-polar meshes. In the following, the coordinates
$(r,\theta,\phi)$ denote spherical radius, co-latitude and azimuth,
respectively. We moreover ignore explicit factors of the magnetic
permeability $\mu_0$ in our notation.  Simulations were run using the
single-fluid \NIII code, which implements a second-order-accurate
Godunov scheme \citep{2004JCoPh.196..393Z}. The code is formulated on
orthogonal curvilinear meshes \citep{2011JCoPh.230.1035Z} and employs
the constrained transport method
\citep{1988ApJ...332..659E} to maintain the divergence-free constraint
of the magnetic field. As an alteration to the publicly available
distribution of the code, we here adopt the upwind reconstruction
technique of \citet{2008JCoPh.227.4123G} to obtain the edge-centered
electric field components needed for the magnetic field update
\citep[also see][]{2010ApJS..188..290S}. This modification became
necessary to take advantage of the more accurate HLLD approximate
Riemann solver of \citet{2005JCoPh.208..315M}, which offers advanced
numerical accuracy for flows in the subsonic regime.

Our numerical model is based on solving the standard resistive MHD
equations including Ohmic resistivity as well as an electromotive
force resulting from the mutual collision of ions and neutrals. Given
the typical number densities within PPDs, the applicability of the
strong-coupling approximation is warranted by detailed chemical
modeling \citep{2011ApJ...739...50B}. Accordingly the gas dynamics can
be modeled by a single-fluid representing the motion of the neutral
species, and the additional term can simply be evolved in a
time-explicit fashion \citep{2009ApJS..181..413C}. The total-energy
formulation of the \NIII code is expressed in conserved variables
$\rho$, $\rho\V$, and $e\equiv \epsilon + \rho\V^2\!/2 + \B^2\!/2$,
and if we define the total pressure, $p^{\star}$, as the sum of the
gas and magnetic pressures, we can write our equation system as
\begin{eqnarray}
  \partial_t\rho +\nabla\cdt(\rho \V) & = & 0             \,, \nonumber\\[4pt]
  \partial_t(\rho\V) +\nabla\cdt
          [\rho\mathbf{vv}+p^{\star}I-\mathbf{BB}] & = &
          - \rho \nabla\Phi                               \,, \nonumber\\[4pt]
  \partial_t e + \nabla\cdt
          [(e + p^{\star})\V - (\V\cdt\B)\B] & = &
          \nabla \cdot \left[\, (\EO\!+\!\EAD)\tms\B \right]  \nonumber\\
      & & - \rho (\nabla\Phi)\cdt\V + \Gamma              \,, \nonumber\\[4pt]
  \partial_t \B -\!\nabla\tms(\, \V\tms\B
                           + \EO + \EAD \,)& = & 0        \,,
\label{eq:mhd}
\end{eqnarray}
where the electromotive forces stemming from the Ohmic and ambipolar
diffusion terms are given by
\begin{equation}
  \EO \equiv - \etao \,(\nabla\tms\B)\,,\quad\textrm{and}
  \label{eq:emf_o}
\end{equation}
\begin{equation}
  \EAD \equiv \etad \,\left[\,(\nabla\tms\B)\times\bb\,\right]\times\bb\,,
  \label{eq:emf_ad}
\end{equation}
(with $\bb\equiv \B/|\B|$ the unit vector along $\B$),
respectively. The gravitational potential $\Phi(r) \equiv -G \Msun/r$
is a simple point-mass potential of a solar-mass star. We moreover
obtain the gas pressure as $p = (\gamma-1)\epsilon$, where
$\gamma=7/5$ is appropriate for a diatomic molecular gas. The source
term $\Gamma$, mimicking radiative heating and cooling, is included to
relax the specific thermal energy density $\epsilon$ to its initial
radial temperature profile. The relaxation is done on a fraction of
the local dynamical time scale, which is short enough to avoid strong
deviations from the equilibrium model but long enough to suppress
vertical corrugation of the disk due to the Goldreich-Schubert-Fricke
instability \citep*[see][for a detailed study of this
  phenomenon]{2013MNRAS.435.2610N}. We remark that, while this
instability is physical in nature, its appearance may be exaggerated
in a strictly isothermal simulation, and more realistic modeling
including radiative transfer should be employed to study its
relevance.

Because we use a time-explicit method, large peak values in the
dissipation coefficients $\etao$ and $\etad$ impose severe constraints
on the numerically permissible integration time step. We address this
by using a state-of-the-art second order super-time-stepping scheme
recently proposed by \citet{2012MNRAS.422.2102M}. In comparison with
conventional first-order methods, this Runge-Kutta-Legendre scheme is
free of adjustable parameters and has been proven superior in terms of
both accuracy and robustness. To maintain the second-order accuracy of
our time integration, we use Strang-splitting for the diffusive terms.

In our previous local simulations \citep[cf. appendix B1
  in][]{2012MNRAS.422.1140G}, we have found that applying a constant
cap on $\etao$ can \emph{qualitatively} change the way in which the
top and bottom disk layers are coupled to each other. For a typical
wind configuration, the horizontal field components generally change
sign at the midplane. We imagine that in this situation a clipped
$\etao$ would affect the amount of field diffusing into the midplane
region and conversely the strength of the current sheets above and
below the magnetically decoupled layer. We therefore choose not to
apply a cap on $\etao$. However, we limit $\etad$ such that
$\Lambda_{\rm AD}\,\beta_{\rm p} \equiv 2\,\Rm\,_{\rm AD}\ge 0.1$
(also see \BS), where $\beta_{\rm p}\equiv 2p/B^2$ is the plasma
parameter, and $\Rm\,_{\rm AD}\equiv \cS\,H/\eta_{\rm AD}$ is the
equivalent of a magnetic Reynolds number for AD. This we find greatly
reduces the numerical cost without noticeably altering the obtained
solution.

\subsection{External ionization and disk chemistry} % ---
\label{sec:ionisation}

The Ohmic and ambipolar diffusivities are evaluated dynamically for
each grid-cell and are based on a precomputed look-up table, that has
been derived self-consistently from a detailed chemical model
accounting for grain-charging and gas-phase chemistry. The ionization
rates that enter this equilibrium chemistry are governed by ionizing
radiation entering the disk. The column densities that govern
  the attenuation of the external ionizing radiation are also computed
  on-the-fly, that is, they reflect the instantaneous state of the
  disc. In principle, this allows for a self-limiting of the emerging
  wind via shielding of ionizing radiation
\citep{2012ApJ...758..100B}.

In addition to some minor modifications to the grain charging
prescription (as detailed below), we have improved the realism of our
ionization model compared to our previous work. The main alterations
concern the inclusion of two additional radiation sources, an
un-scattered direct X-ray component, and hard UV photons. These have
in common a shallow penetration depth but comparatively high flux
levels, thus mainly affecting the ionization level of the disk's
surface layers.

\subsubsection{FUV ionization layer}
\label{sec:fuv}

Adopting a very simple prescription based on the recent study by
\citet{2011ApJ...735....8P}, we have added the effect of an FUV
ionization layer based on an assumed ionization fraction of
$f=2\times10^{-5}$ (representing full ionization of the gas-phase
carbon and sulfur atoms susceptible to losing electrons when struck
by FUV photons), and a collision rate coefficient of $2\times10^{-9}
\m^3\s^{-1}$. We further assume that FUV photons penetrate to a gas
column density of $0.03\g\,\cm^{-2}$. Due to their lower amplitude, we
ignore any scattered, diffuse or ambient FUV illumination of the disk
and only evaluate sight-lines pointing directly towards the central
star. Note that this deviates from the local simulations of
\BS, where the vertical gas column was used
for attenuation of the incident flux. Our treatment is motivated by
the assumption that a large fraction of the FUV photons originate
directly from the central star \citep[see upper right panel of figure
  9 in][who study the forward scattering of FUV photons in
  detail]{2011ApJ...739...78B}.

\subsubsection{Illumination by X-rays}

A similar treatment is adopted for the unscattered X-ray component,
for which we use the fit to the Monte-Carlo radiative-transfer results
of \citet{1999ApJ...518..848I} given by eq.~(21) in
\citet{2009ApJ...701..737B}. Deviating from our previous local
simulations \citep[cf. eq (1) in][]{2011MNRAS.415.3291G}, our new
X-ray illumination is
\begin{eqnarray}
  \zeta_{\rm XR} & = &
  10^{-15}\s^{-1}\,
  \left[ \exp{-\left(\frac{\Sigma_A}{\Sigma_{\rm sc}}\right)^{\!\!\alpha}}
       + \exp{-\left(\frac{\Sigma_B}{\Sigma_{\rm sc}}\right)^{\!\!\alpha}}
  \;\right]\;{\it r}^{-2}\nonumber \\[2pt]
  & + & 6\tms10^{-12}\s^{-1}\;\;
  \exp{\,-\left(\frac{\Sigma_C}{\Sigma_{\rm ab}}\right)^{\,\beta}}\;{\it r}^{-2}
  \label{eq:xray}
\end{eqnarray}
where $\Sigma_A$ ($\Sigma_B$) is the gas column to the top (bottom) of
the disk surface, and $\Sigma_C$ is the column density along radial
rays towards the star. The radial column contains a contribution from
the inner disk (which is not part of the computational domain) and
reaches in to an inner truncation radius of $r_{\rm tr}=5 R_\odot$.
Adopting values from \citet{2009ApJ...701..737B}, the shape
coefficients are $\Sigma_{\rm sc}=7\tms10^{23}\cm^{-2}$, and
$\alpha=0.65$ for scattered X-rays, and $\Sigma_{\rm
  ab}=1.5\tms10^{21}\cm^{-2}$, and $\beta=0.4$, for absorption of the
direct component, respectively. Both contributions additionally decay
with the squared radius to account for the decrease in X-ray
luminosity. To account for typical median values of
$10^{30}\erg\s^{-1}$ in observational data of luminosities in young
star clusters such as in the Orion Nebula \citep{2000AJ....120.1426G},
we enhance the incident X-ray flux by a factor of $5\times$ compared
to the above stated values. In view of future work, we envisage
additional improvements to the X-ray ionization model by adopting the
new results of \citet{2013MNRAS.436.3446E}, who consider the enhanced
ionizing effects of X-rays due to the presence of heavier elements
(assuming solar abundance). These alterations have been found to lead
to enhanced ionizing fluxes at intermediate column densities compared
to the original results of \citet{1999ApJ...518..848I}.

\subsubsection{Modifications to the disk chemistry}

Our chemical network is based on the one used by
\citet{2006A&A...445..205I}, labeled \texttt{model4}, with grain
surface reactions and a simplified gas-phase reaction set involving
one representative metal and one molecular ion. Small changes in
recent years include correcting the electron sticking probabilities
for the grain charge \citep{2011ApJ...739...50B} and consistently
treating the molecular ion such as HCO$^+$ (with a mass of 29 protons)
since this is the long-lived species in a series of several
reactions. Further details on the reaction network and diffusivities
can be found in section 2.2. of \citet{2013ApJ...771...80L} as well as
section 4.2 of \citet{2013ApJ...764...65M}.

\begin{figure}
  \center\includegraphics[width=\columnwidth]{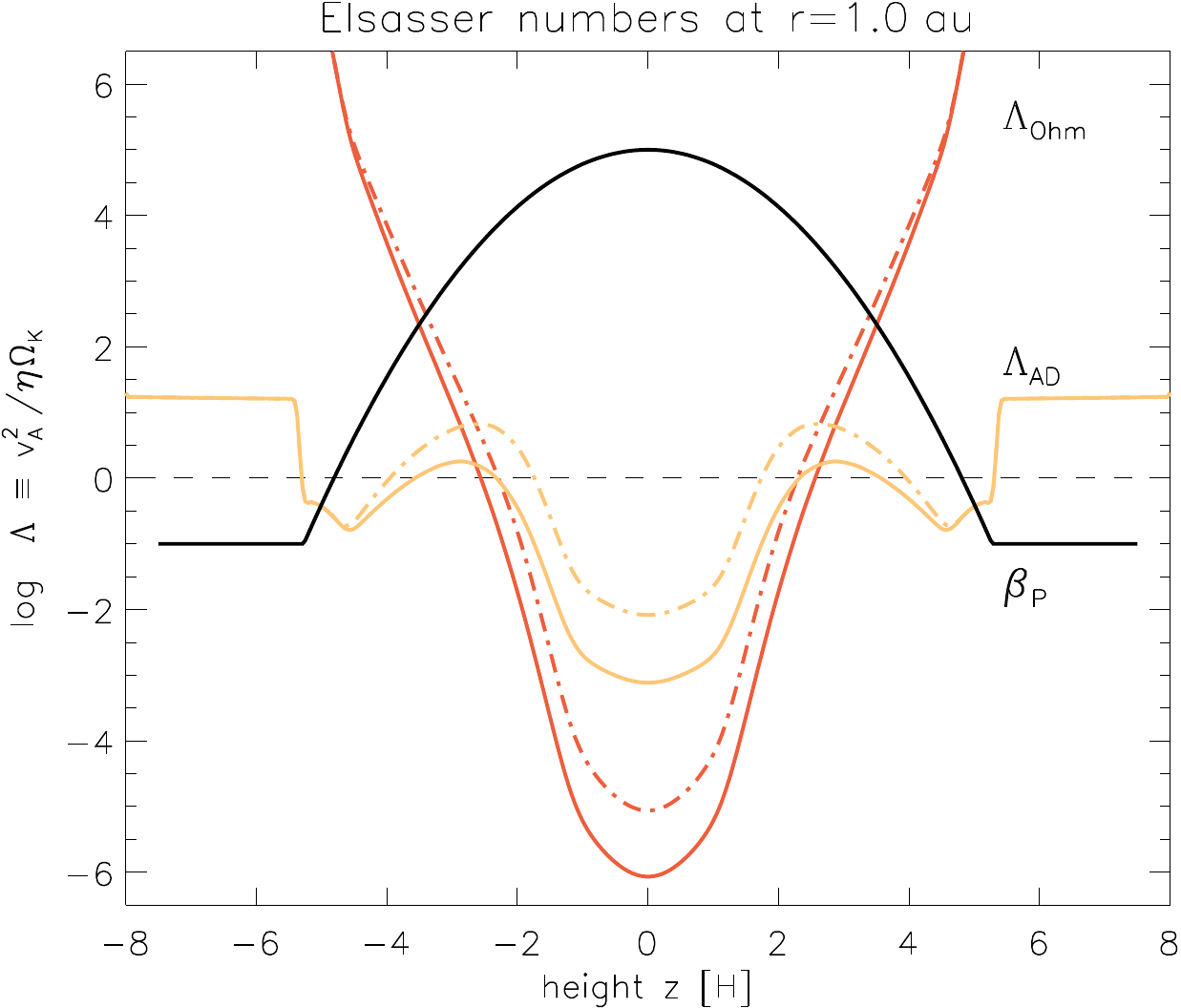}
  \caption{Profiles of the initial Elsasser numbers (at radius
    $r=1\au$) for ohmic resistivity (light orange), and ambipolar
    diffusion (red). Solid lines represent the fiducial case of a
    dust-to-gas mass ratio of $10^{-3}$, whereas dash-dotted lines
    show profiles for a value of $10^{-4}$. We also plot the initial
    profile of the plasma $\beta$ parameter (solid black line).}\medskip
  \label{fig:elsa}
\end{figure}

The resulting \emph{initial} ionization profile expressed in Elsasser
numbers $\Lambda_{\rm O/AD} \equiv v_{{\rm A}\,z}^2\, (\Omega\,
\eta_{\rm O/AD})^{-1}$ with $v_{{\rm A}\,z}\equiv B_z/\!\sqrt{\rho}$
is plotted in Fig.~\ref{fig:elsa} at a radius of $1\au$, together with
the height-dependent plasma $\beta$ parameter. The implications of
this figure for our simulations are discussed in
Sect~\ref{sec:prelim}. The region with $|z|\simgt 5\,H$, where the
plasma parameter becomes constant with height is caused by applying a
lower limit on the gas density to avoid excessive Alfv{\'e}n speeds in
the disk halo. Following \BS, we chose this
limit at $10^{-6}$ (in our global model, this value is with respect to
the initial midplane value at any given cylindrical radius). We remark
that the equilibrium density profile is artificially steep in our
isothermal disk model with temperature constant on cylinders. In this
respect, the density floor acts to mimic a disk with realistic
radiation thermodynamics, where the irradiation heating of the upper
disk layers naturally creates an increased pressure scale height and
hence a much shallower density profile. Because $\eta_{\rm AD}$ (and
hence $\Lambda_{\rm AD}$) depends on the magnetization of the plasma,
the actual density profile in the disk halo has an influence on how
well the gas is coupled to the magnetic field lines, which in turn
affects the efficiency of the magneto-centrifugal wind launching
mechanism. This caveat should be kept in mind when interpreting the
mass loading of the wind, which we expect to be a function of the
thermodynamics in real protoplanetary disks.

\subsection{Global disk model} % ---
\label{sec:disk_model}

We now describe the underlying protostellar disk model, which is
largely identical to the one used in \citet{2013ApJ...779...59G}.  The
equilibrium disk model is based on a locally-isothermal temperature,
$T$, decreasing with cylindrical radius as $T(\Rc) = T_0\,
(\Rc/\Rc_0)^q$. For $q=-1$, such a profile leads to a constant opening
angle throughout the disk as it is commonly used for global numerical
simulations of PPDs, and which provides us with the reference disk
model here. To study what effect disk flaring has on the absorption of
direct FUV photons, we also produce a moderately flaring disk with
$q=-3/4$. For this value, the resulting power-law index for the local
opening angle, $h(R)\equiv H(R)/R$, is $\psi \equiv (q+1)/2= 0.125$,
which is in agreement with observational constraints for this value
based on sub-mm observations \citep{2009ApJ...700.1502A}, and from
multi-wavelength imaging \citep{2008A&A...489..633P}.\footnote{At
  least when assuming well-mixed dust grains, since both observations
  are sensing the dust content of the disk rather than the gas
  density.} The radial power-law exponents for the gas surface
density, $\Sigma(\Rc)$, are $-1/2$ for our fiducial model, and $-3/8$
for the flaring disk model, respectively.

Our temperature distribution is complemented by a power-law for the
midplane density, $\rho_{\rm mid}(\Rc) = \rho_0\, (\Rc/\Rc_0)^{p}$,
with $p=-3/2$. The equilibrium disk solution is given by
\begin{equation}
  \rho(\rr) = \rho_0 \left( \frac{\Rc}{\Rc_0} \right)^{p}
                      \exp{\left(\,\frac{G \Msun}{\cS^2}
                        \left[ \frac{1}{r} - \frac{1}{\Rc}
                          \right]\,\right)}\,,\;\textrm{and}\quad
  \label{eqn:rho}
\end{equation}
\begin{equation}
  \Omega(\rr) = \Omega_{\rm K}(\Rc)\,
  \left[\,(p+q) \left(\frac{H}{\Rc}\right)^{\!\!2}
    \!+ (1+q) - \frac{q \Rc}{r}\,\right]^\frac{1}{2}\!\!,
  \label{eqn:Omega}
\end{equation}
which can be derived from the requirement of hydrostatic/-dynamic
force balance in the vertical and radial directions, and where the
Keplerian angular velocity $\Omega_{\rm K}(\Rc) = \sqrt{G\Msun}\,
\Rc^{-3/2}$. In the above equations, the square of the isothermal
sound speed results from the prescribed temperature profile as
$\cS^2=c_{\rm s0}^2\, (\Rc/\Rc_0)^q$, with a parameter $c_{\rm s0}$
corresponding to a value of $ H(\Rc) \equiv c_{\rm s} \Omega_{\rm K}^{-1}
=0.05\Rc$.

Guided by previous results of local 3D simulations that exhibit
quasi-axisymmetric structure, we mainly focus on 2D-axisymmetric
simulations.  Our computational domain covers
$r\in[0.5,5.5]\au$. In the latitudinal direction, the grid covers eight
pressure scale heights on each side of the disk midplane, that is,
$\theta\in[\pi/2-\vartheta,\pi/2+\vartheta]$, with
$\vartheta=8\,H/r$. In the case of the flaring disk, the coverage of
scale heights varies from $8\,H$ at $r=1\au$ to $6.5\,H$ at $r=5\au$.
The grid resolution is $N_r \times N_\theta \times N_\phi = 512 \times
192 \times 64$, which means that we are able to resolve relevant MRI
wave lengths. In the axisymmetric models we use $N_r \times N_\theta =
1024 \times 384$ cells, corresponding to $\ge 24$ grid points per
pressure scale height, matching the resolution of the
three-dimensional box simulations of \BS.  For the two
non-axisymmetric simulations, the azimuthal extent is restricted to
$\phi\in[0,\varphi]$, with $\varphi=\pi/2$ (a quarter wedge), or
$\pi/4$ to limit the computational expense. We furthermore note that
in the ideal-MHD case, where turbulence develops, the azimuthal domain
size has been found to have an effect on the final saturated
state\citep{2011MNRAS.416..361B,2012ApJ...744..144F,2012ApJ...749..189S},
at least for the case of zero net flux.

Because our ionization model depends on real physical values of the
gas column density, we have to chose a meaningful normalization factor
$\rho_0$, which we fix according to a surface density of
$\Sigma=150\g\cm^{-2}$ at the location $\Rc=5\au$, compatible with the
typical minimum-mass solar nebula
\citep[MMNS][]{1981PThPS..70...35H}. In physical units, the disk
temperature is $T=540\K$ at a radius of $\Rc=1\au$, and $T=108\K$ at
$5\au$, resembling typical expected conditions within the protosolar
disk. While our values for $\rho$, $T$ and $\Sigma$ are similar to the
MMSN values at $5\au$, our adoption of different values for the radial
power-law indices means that these values are different at $1\au$
compared to those adopted by \BS.

Our model disk is initially threaded by a weak vertical magnetic field
$B_z(\Rc)$ corresponding to a midplane $\beta_{\rm p\,0}\equiv
2p/B^2=10^5$ independent of radius for our fiducial disk. To achieve
this, the vertical magnetic flux has to vary as a power-law in radius,
taking into account the radial distribution of the midplane gas
pressure that itself depends on the density and temperature. In our
disk model, the gas pressure, $p(\Rc)$ decreases outward, implying
that $B_z(\Rc)$ falls off with radius, too. While such a configuration
is generally expected when accounting for inward advection and outward
diffusion of the embedded vertical flux \citep{2014ApJ...785..127O},
our particular choice of keeping the \emph{relative} field strength
constant is of course arbitrary.  To preserve the solenoidal
constraint to machine accuracy, we compute the poloidal field
components from a suitably defined vector potential
$A_\phi(r,\theta)$. The initial disk model is perturbed with random
white-noise fluctuations in the three velocity components. The rms
amplitude of the perturbation is chosen at one percent of the local
sound speed. We furthermore add a white-noise perturbation in the
magnetic field with an rms amplitude of a few $\mu{\rm G}$, which is
on the sub-percent level compared with the net-vertical field of
typically a few ten $\mG$.

\subsection{Boundary conditions} % ---
\label{sec:bcs}

To complete the description of our numerical setup, we need to specify
boundary conditions (BCs). For the fluid variables, we employ standard
``outflow'' conditions at the inner and outer radial domain edges,
$r_{\rm in}$ and $r_{\rm out}$. This type is a combination of a
zero-gradient condition in the case of $\,\hat{\bf n}\cdot\V(r_{\rm
  in},\theta)>0\,$ (where $\hat{\bf n}$ denotes the outward-pointing
normal vector), and reflective BCs in the opposite case, thus
preventing inflow of material from outside the domain. At the upper
and lower boundaries (that is, in the $\theta$ direction), we
furthermore balance the ghost zone values such that a hydrostatic
equilibrium is obtained. This helps to control artificial jumps of the
fluid variables adjacent to the domain boundary as they are frequently
encountered with finite-volume schemes.

In contrast to our earlier global simulations of layered accretion
disks subject to Ohmic resistivity and containing an embedded
gas-giant planet \citep{2013ApJ...779...59G}, we here make a different
choice for the vertical magnetic-field boundary condition. Whereas we
previously applied magnetic BCs of the \emph{perfect conductor} type
(that is, forcing the normal field component to zero and allowing
non-zero parallel field), we here use \emph{pseudo vacuum} conditions,
conversely enforcing the perturbed parallel field to vanish at the
surface and only allowing a perpendicular perturbed field.  Note that
we exclude the initial net-vertical field from the procedure such that
only deviations from the background field are subject to the
normal-field condition.  The vertical flux threading the disk is thus
preserved.  Since the disk's upper layers are magnetically dominated,
and the Lorentz force acts to align the flow lines with the magnetic
field, letting the field lines cross the domain boundary is of course
a prerequisite for launching an outflow.  While we realize that
forcing the radial and azimuthal components of the perturbed field to
vanish at the boundary may unduly restrict the magnetic topology of
the emerging wind solution, we postpone the study of less-restrictive
but more cumbersome boundaries to future work.

Unlike in a radially-periodic local shearing-box simulation, our
global model is critically affected by the inner radial boundary
condition. In a real protoplanetary disk, we can expect that alkali
metals will be thermally ionized in this region, leading to the
development of the MRI on timescales short compared with the orbital
timescale at the inner domain boundary of our model. This poses the
question how to best interface the MRI-turbulent inner disk with the
magnetically decoupled midplane of the outer disk that we model. It is
likely that MRI-generated fields can efficiently leak into the outer
disk via magnetic diffusion. In a first attempt to account for the
MRI-active inner disk, we gradually taper the diffusivity coefficients
to zero within the inner $0.5\au$ of our disk model. Studying the
inner edge of the dead zone will however require dedicated simulations
\citep[similar to the ones performed by][]{2010A&A...515A..70D}
including this transition region.

% ------------------------------------------------------------------------------

\section{Results}
\label{sec:res}

The main aim of our paper is to establish the laminar wind solutions
that \BS previously found in local shearing
box simulations in the context of global disk simulations. In the
interest of economic use of computational resources and to guarantee
adequate numerical resolution of our global models, we primarily focus
on 2D axisymmetric simulations, but have also run three-dimensional
simulations to check for non-axisymmetric solutions.  Since all our
runs include a net-vertical flux, axisymmetry is warranted to obtain a
reading on the development of the MRI.  In cases where the solution
proves laminar, axisymmetry is likely to produce a reasonably accurate
picture.  We address the possibility of non-axisymmetric secondary
instabilities using a limited set of three-dimensional calculations,
described in Section~\ref{sec:3d}.

% ---

\begin{table}\begin{center}
\caption{Summary of simulation runs.
  \label{tab:models}}
\begin{tabular}{lccccccc}\hline
 Run label & $\!$Ohm$\!$ & AD & $\beta_{\rm p\,0}$ & d/g & $q$ & Resol.\\ 
\hline
 O-b6      & \yes & \no  & $10^6$ & $10^{-3}$ & -1   & $1024\tms 384$\\[4pt]
 OA-b5     & \yes & \yes & $10^5$ & $10^{-3}$ & -1   & $1024\tms 384$\\
 OA-b6     & \yes & \yes & $10^6$ & $10^{-3}$ & -1   & $1024\tms 384$\\
 OA-b7     & \yes & \yes & $10^7$ & $10^{-3}$ & -1   & $1024\tms 384$\\[4pt]
 OA-b5-d4  & \yes & \yes & $10^5$ & $10^{-4}$ & -1   & $1024\tms 384$\\
 OA-b5-flr & \yes & \yes & $10^5$ & $10^{-3}$ & -3/4 & $1024\tms 384$\\
 OA-b5-flr-nx & \yes & \yes & $10^5$ & $10^{-3}$ & -3/4 & $512\tms 192\tms 128$\\
 OA-b5-nx  & \yes & \yes & $10^5$ & $10^{-3}$ & -1   & $512\tms 192\tms 64$\\
 \hline
\end{tabular}
\end{center} 
\footnotesize Models are labeled according to the included
micro-physical effects (first two letters) and the strength of the
net-vertical magnetic field (expressed in terms of the midplane value
$\beta_{\rm p\,0}$, prefixed with the letter `b'). Further labels
refer to the dust-to-gas mass ratio (`d/g', prefixed with the
  letter `d'), the power-law index, $q$, of the radial temperature
profile (`flr' for ``flaring''), and whether the azimuthal
  dimension is included (`nx' for ``non-axisymmetric'').
\end{table}

% ---

The simulations conducted for this work are listed in
Table~\ref{tab:models}, where we summarize key model parameters. We
assume a fiducial dust-to-gas mass ratio of $10^{-3}$, that is, a
depletion by a factor of ten compared to interstellar abundances.
Starting from the classic layered PPD (model `O-b6') including only
Ohmic diffusivity, we compare this standard ``dead-zone'' disk with a
corresponding model, `OA-b6', additionally including the effects of
ambipolar diffusion. Our fiducial reference model is `OA-b5', with a
midplane plasma parameter of $10^5$. In the presence of combined
ambipolar diffusion and Ohmic resistivity, we observe the waning of
the MRI, which is instead replaced by a laminar wind solution. Unlike
in geometrically constrained shearing box simulations
\citep[\BS,][]{2013ApJ...772...96B}, we naturally obtain a
field configuration with field lines bending outward on both
``hemispheres'' of the disk. Notably, this topology produces thin
current layers, which have previously been discussed by
\BS. The stability and evolution of these
current layers will be one focus of our paper.

We perform additional analysis on the influence of further key input
parameters. With the global disk model allowing direct illumination
from the central star, it becomes possible to address the question of
how the disk's ionization is affected by disk flaring. This is studied
in model `OA-b5-flr', where we use a power-law index of $q=-3/4$
instead of $q=-1$ to obtain a moderately flaring disk surface as is
supported by observations
\citep{2008A&A...489..633P,2009ApJ...700.1502A}. The role of dust
depletion, driven by processes such as coagulation into larger grains
and/or sedimentation, is considered in model `OA-b5-d4' with a reduced
dust-to-gas mass fraction of $10^{-4}$ (as used in \BS) compared with
our fiducial value of $10^{-3}$. For the sake of brevity, we here
refrain from varying any of the many other input parameters of our
model, as for instance, the X-ray or CR intensities, or the FUV
penetration depth, as well as parameters affecting the disk
thermodynamics.

\subsection{Preliminaries} % ---
\label{sec:prelim}

The inputs to our models are very similar to those included in \BS, so
it is instructive to compare the Elsasser number profiles in our model
with their fiducial model as a means of understanding the similarities
and differences between their results and ours.  The fiducial model
presented in \BS is computed at $1\au$, and has a dust-to-gas mass
fraction of $10^{-4}$, so we can compare this with the dot-dashed
lines in Figure~\ref{fig:elsa}. The profiles shown there, and in
figure~1 of \BS are similar, with $\beta_{\rm P}$ decreasing below
unity at disk heights $|z| \gtrsim 4.5H$, preventing MRI turbulence
from developing at these high altitudes.  In their initial condition
and ours, $\Lambda_{\rm O}$ increases monotonically from the midplane
upward.  The two initial states also have similar profiles for
$\Lambda_{\rm AD}$, displaying local maxima at intermediate disk
heights ($z \sim \pm 2.5 H$). The differences in the surface densities
in the \BS models and ours at $1\au$, combined with our inclusion of a
direct X-ray component in addition to the scattered component means
that $\Lambda_{\rm AD} \sim 10$ at this local maximum in our model,
whereas it only rises modestly above unity in \BS. We therefore expect
that our model with a dust-to-gas mass fraction of $10^{-4}$ should
contain narrow regions at intermediate heights that support the growth
of MRI channel modes. It is noteworthy that \BS also observed the
development of the MRI at the beginning of their simulations, but
report that the subsequent amplification of the field causes the
ambipolar diffusion to increase. Their disk then relaxes to completely
laminar state in which angular momentum transport is driven by a
magneto-centrifugal wind. The larger value of $\Lambda_{\rm AD}$ in
our models may allow MRI turbulence to be sustained in these regions,
or may instead lead to quenching of the MRI after it has reached a
more nonlinear stage of development. In this paper, we define our
fiducial model to be one in which the dust-to-gas mass fraction is
$10^{-3}$, and the Elsasser numbers for this case are shown by the
solid lines in Fig.~\ref{fig:elsa}.  The larger dust concentration
reduces the gas-phase electron fraction and Elsasser numbers, and the
local maximum in $\Lambda_{\rm AD}$ now peaks moderately above unity
(but still attaining a larger peak value than the fiducial model in
\BS).

\begin{figure*}
  \center\includegraphics[width=1.5\columnwidth]{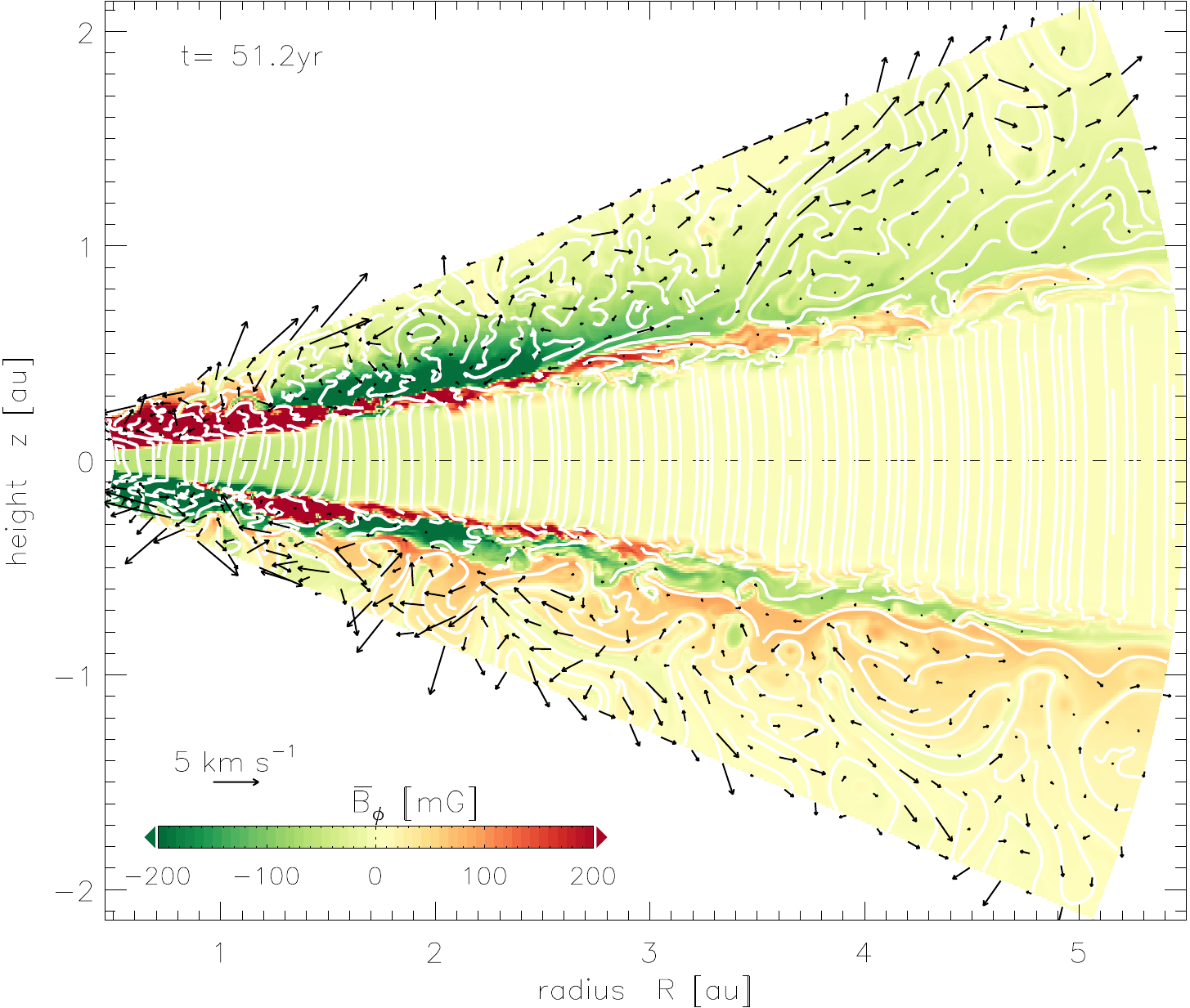}
  \caption{Visualization of the global disk structure for model O-b6
    with only Ohmic resistivity and $\beta_{\rm p\,0}=10^6$. The
    azimuthal fields reach peak strengths of $\simeq 2.5\,{\rm G}$ in
    the active disk layers. Vertical outflow is intermittent, but
    regions of an ordered wind geometry can be identified (see upper
    half around $3.5 - 4\au$).}\bigskip
  \label{fig:vis2D_O-b6}
\end{figure*}

\subsection{Traditional dead zone model} % ---
\label{sec:deadzone}

We begin discussion of the simulation results by considering the
2D axisymmetric run
O-b6, for which Ohmic resistivity is included and ambipolar diffusion
is neglected. To introduce our setup, and give the reader an
impression of the general appearance of our PPD model,
Fig.~\ref{fig:vis2D_O-b6} visualizes the magnetic field and poloidal
flow vectors from the model. The color coding of the azimuthal field
strength shows the MRI-turbulent surface layers with tangled poloidal
field lines superimposed in white.  In the outer disk, where the MRI
has not fully developed yet, the parity of the horizontal field is odd
with respect to reflection at the $z=0$ line, consistent with the
winding-up of vertical field, $B_z$, by the weak differential rotation
$\partial_z\,\Omega$ of our disk model. While the inner domain
boundary also shows an odd field symmetry, intermediate sections of
the disk show a mixed parity, reflecting the presence of MRI modes
with both even and odd vertical wave numbers. Quite remarkably, the
MRI channels extend over several AU in the radial direction, although
we remark that this may be overly pronounced in the axisymmetric case,
where the growth of parasitic modes is likely to be artificially
restricted \citep{2008ApJ...684..498P}.  Because we use a net-vertical
field, the linear MRI modes appear more pronounced than in the
otherwise comparable 3D simulations of \citet{2010A&A...515A..70D}
without a net $B_z$ field.

The poloidal velocity field plotted as black arrows is mostly
disordered in the turbulent regions but also shows some level of
coherence in the upper layers, where a wind topology can be
seen. While the wind is intermittent rather than steady and laminar, a
general tendency for outflow is seen. This is consistent with similar
observations of disk winds being launched from fully-MRI-active
accretion disks \citep[see,
  e.g.,][]{2009ApJ...691L..49S,2013A&A...552A..71F}. We quantify the
mass loss via the vertical disk surfaces by evaluating
\begin{equation}
  \dot{M}_{\rm wind} \equiv 
  \left.  2\pi\,\int_{r=r_{\rm i}}^{r_{\rm
      o}}\, \rho v_{\theta}\,r\,\sin\theta\,{\rm d}r 
  \;\right|_{\theta=\theta_1}^{\theta_2}
  \label{eq:mdot_wind}
\end{equation}
at the upper and lower disk surface, $\theta=\theta_1,\theta_2$,
respectively. For model O-b6, we find a net mass loss rate of
$\dot{M}_{\rm wind}=1.47\pm 0.37 \times 10^{-8}\Msun\yr^{-1}$ (also
cf. Table~\ref{tab:results} below).

We observe the formation of field arcs and material flowing downward
along field lines towards the arc foot points. This can be seen in the
lower disk half around $r=2.8\au$, and $r=4.3\au$, for example, and
illustrates the potential role of buoyancy instability in regulating
the disk's magnetization. Amplification of the magnetic field through
the growth of channel modes, and their break up through the action of
parasites, apparently leads to the formation of these locally uprising
field structures. As an alternative explanation, we remark that the
dynamic evolution of such magnetic loops has previously been
attributed to the effect of differential rotation onto the magnetic
footpoints \citep{1998ApJ...500..703R}. 

In contrast, within the magnetically decoupled midplane layer, the
magnetic field remains predominantly vertical, retaining the initial
field configuration.  Near the dead-zone edge, the effect of Ohmic
diffusion gradually declines. There, the azimuthal magnetic field is
allowed to change its value, resulting in localized current
layers. This is very similar to the situation obtained in the disks
that include ambipolar diffusion, as discussed later.

\begin{figure}
  \center\includegraphics[height=\columnwidth]{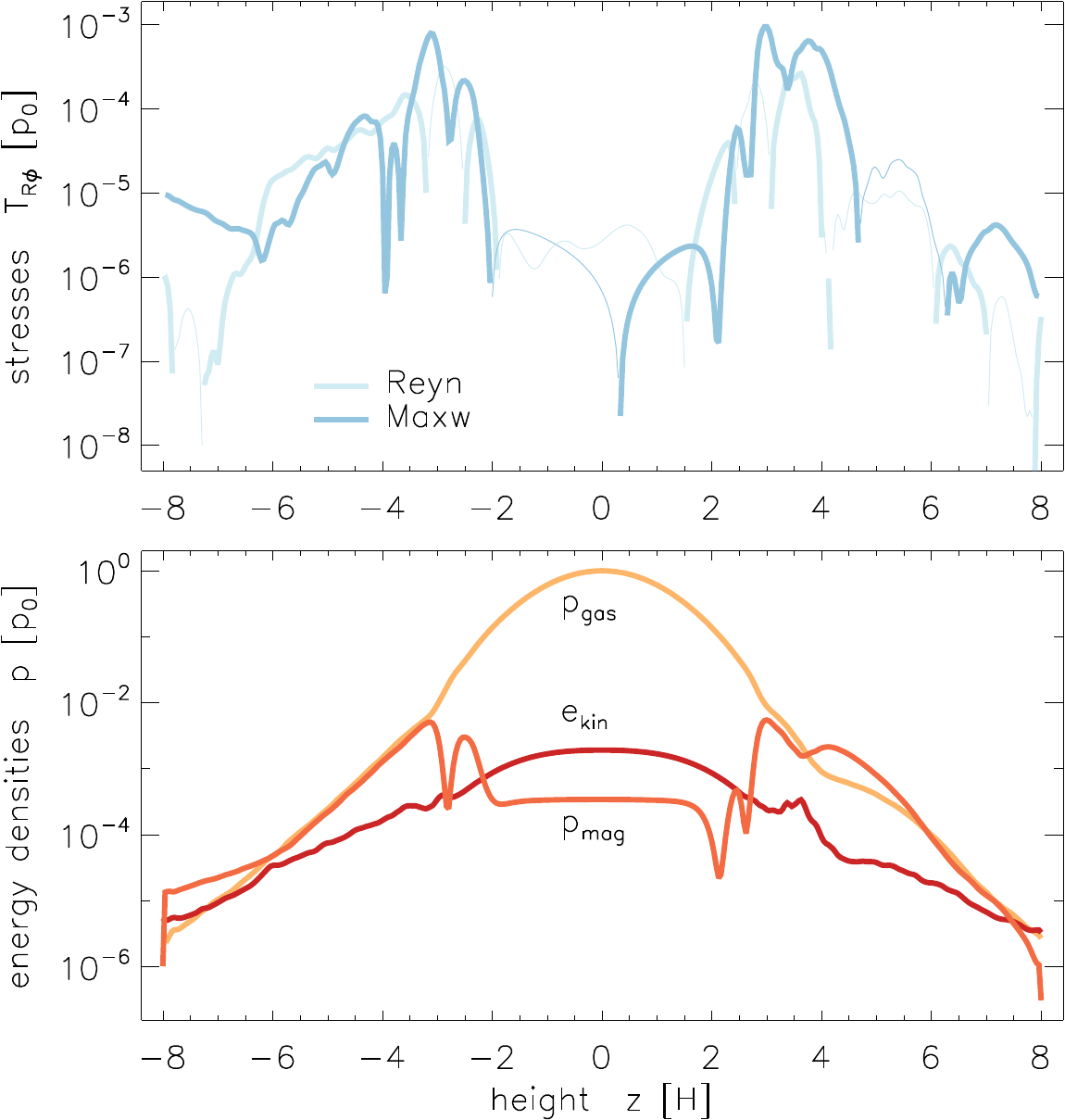}
  \caption{Vertical profiles averaged over $r\in [2,3]\au$ and $t=25$
    orbits for model O-b6 with Ohmic resistivity only and a midplane
    $\beta_{\rm p\,0}=10^6$. Negative values of the stresses (upper
    panel) are represented by thin lines.}
  \label{fig:prof_O-b6}
\end{figure}

We illustrate the vertical disk structure in Fig.~\ref{fig:prof_O-b6},
where we plot, in the upper panel, the total Reynolds and Maxwell
stresses relative to the \emph{initial} midplane gas pressure,
$p_0$. The profiles show the typical signature of a laminar dead zone
and MRI-turbulent surface layers
\citep{2003ApJ...585..908F,2007ApJ...670..805O,%
  2011MNRAS.415.3291G,2012MNRAS.422.1140G}, where the mechanical and
magnetic shear-stresses lead to transport of angular momentum. Because
of the relatively weak net-vertical magnetic flux, the level of
turbulence is modest, and the vertically-integrated dimensionless
Maxwell stress is $\simeq 8\times 10^{-5}$. With the ``viscous'' mass
accretion rate (i.e., that effected by internal stresses) approximated
by
\begin{equation}
  \dot{M}_{\rm visc} \approx
  \left.  3\pi\,\Omega^{-1}\!\int_{\theta=\theta_1}^{\theta_2}\!
  \left( T_{R\phi}^{\rm Reyn}+T_{R\phi}^{\rm Maxw} \right)\,r\,{\rm d}\theta
  \;\right|_{r=r_{\rm ref}}\!\!,
  \label{eq:mdot_visc}
\end{equation}
we estimate a value of $\dot{M}_{\rm visc} \simeq 0.3\times
10^{-8}\Msun\yr^{-1}$, at $r_{\rm ref}=2.5\au$, if mass accretion were
effected by turbulent transport of angular momentum alone. This can be
compared to the actual mass accretion rate, which we can easily
compute in our global model via
\begin{equation}
  \dot{M}_{\rm accr} \equiv 
  \left. 2\pi\,\int_{\theta=\theta_1}^{\theta_2}\, 
  \rho v_r\,r^2 \sin\theta\;{\rm d}\theta
  \;\right|_{r=r_{\rm ref}}\;.
  \label{eq:mdot_accr}
\end{equation}
Applying a radial average over $r\in[2,3]\au$, and estimating
fluctuations occurring within a time interval spanning
$t\in[75,100]\yr$, we find $\dot{M}_{\rm accr}=(0.14\pm 1.53)\times
10^{-8}\Msun\yr^{-1}$ (also see the last column of
Table~\ref{tab:results}). Clearly, this diagnostic turns out to be
subject to strong fluctuations for model O-b6 -- presumably due to the
presence of strong radial motion within the MRI channel
modes. Equation~(\ref{eq:mdot_accr}) will however prove useful when
evaluating the radial drift of material in the localized current
layers seen in the AD-dominated disk models. We note at this point
that a simple energy conservation argument prevents the steady-state
mass loss rate (to infinity) through a magneto-centrifugal wind being
larger than the accretion rate through the disk. The large value of
$\dot{M}_{\rm wind}$ reported above relative to $\dot{M}_{\rm accr}$
suggests that the wind loss rate is not yet converged. Indeed we note
that \BS report that increasing the vertical sizes of their shearing
boxes leads to a reduction of the wind mass loss rates. A similar
conclusion is reached by \citet{2013A&A...552A..71F} for winds
launched from turbulent disks. It appears that obtaining accurate and
physically meaningful estimates for the mass fluxes through the upper
boundaries of the disk will require simulations that are truly global
in the vertical direction, such that the fast magnetosonic point is
contained within the simulation domain rather than outside of the
boundary as it is at present for all of the models that we present in
this paper (ensuring that the wind launching region is causally
disconnected from the imposed boundary conditions).

In the lower panel of Fig.~\ref{fig:prof_O-b6}, the gas pressure,
kinetic energy and magnetic pressure, are shown. Within the dead-zone
layer, between $z=\pm 4\,H$, the disk is supported by gas pressure
alone, which follows a Gaussian profile. Owing to the additional
magnetic pressure support within the active layer, the effective scale
height increases roughly twofold there. Note that unlike seen in the
isothermal simulations of \BS, cf. their fig.~3, the magnetic pressure
does not significantly exceed the value of the gas pressure in large
parts of the disk corona. We attribute this difference to the
dissipation heating \citep{2011ApJ...732L..30H} present in our
simulations, which we note, however, may be overly pronounced in the
axisymmetric models. The kinetic energy only becomes important very
close to the vertical boundaries of our model. This is also reflected
in the position of the Alfv{\'e}n point, which is relatively poorly
constrained and which we infer as $(7.60\pm 0.45)\,H$. Values for
these vertical points are listed in Table~\ref{tab:results} for all
models. We report time- and space-averaged values obtained for
$r\in[1,5]\au$, and we have appropriately taken into account the
flaring of the disk. Because of the turbulent character of the upper
disk layers, we cannot obtain a good estimate for the location of the
base of the wind in model O-b6.

\subsection{Laminar disk models with surface winds} % ---

We now discuss models that include both Ohmic resistivity and
ambipolar diffusion.  A key finding of \BS is that the inclusion of
ambipolar diffusion inhibits linear growth of the MRI in the
intermediate disk layers, that is, the regions that are not affected
by Ohmic resistivity. Since the ambipolar diffusion coefficient
depends on the plasma parameter, one might expect that the effect of
AD is less severe for high values of $\beta_{\rm p}$. To test this, we
have run simulations OA-b6, and OA-b7, with $\beta_{\rm p\,0}=10^6$,
and $10^7$, respectively.

\subsubsection{Ambipolar diffusion with weak vertical fields}
\label{sec:ambipolar}

In accordance with the $\beta_{\rm p\,0}=\infty$ simulations of \BS,
we find very low levels of turbulence in these simulations
(cf. Table~\ref{tab:results}). As a consequence, the radial mass
transport via accretion (see $\dot{M}_{\rm accr}$ values) is found to
be negligible. At the same time, because of the weak vertical field,
the magneto-centrifugal wind is absent in model OA-b7, and only poorly
developed (and intermittent) in model OA-b6. We conclude that the
corresponding field amplitude of about $10\mG$ (at $1\au$) can be
assumed as a minimal requirement for significant mass transport by
magnetic effects. In the following, we accordingly focus on models
derived from our fiducial run OA-b5, with $\beta_{\rm p\,0}=10^5$,
that is, $B_{z\,0}=31.5\mG$ ($4.2\mG$) at $1\au$ ($5\au$).

\subsubsection{The fiducial model}
\label{sec:fiducial}

\begin{figure}
  \center\includegraphics[height=\columnwidth]{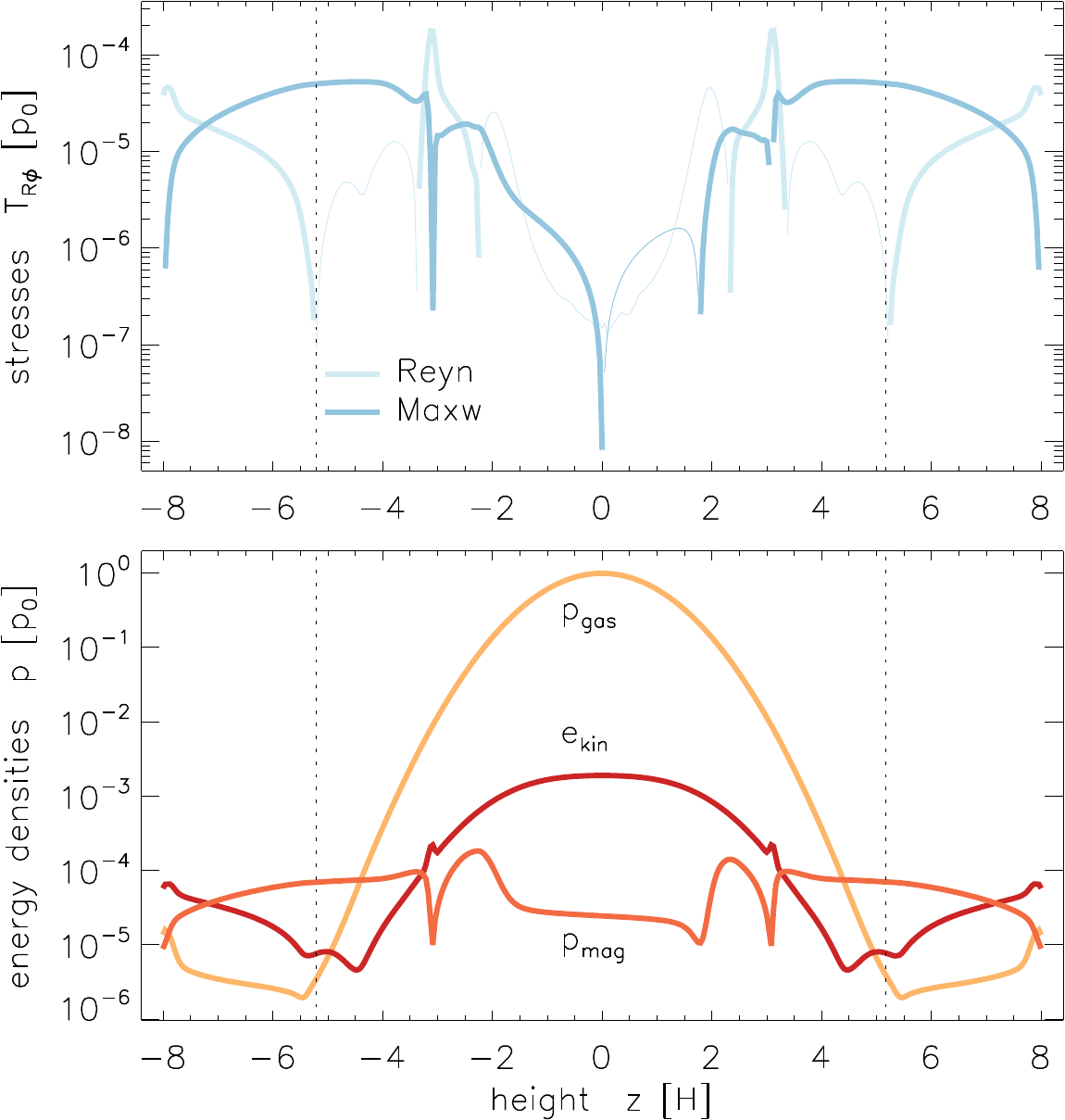}
  \caption{Same as Fig.~\ref{fig:prof_O-b6}, but for model OA-b5
    including both Ohmic resistivity and ambipolar diffusion, and with
    a stronger net-vertical field. Note the different scale of the
    abscissa in the upper panel, reflecting the reduced level of
    ``viscous'' transport. Dotted lines indicate the wind base at $z=\pm
    z_{\rm b}$.}
  \label{fig:prof_OA-b5}
\end{figure}

In Figure~\ref{fig:prof_OA-b5}, we show vertical profiles of the
$R\phi$ stress components for our fiducial model, OA-b5, where owing
to the stronger net vertical field the MRI is ultimately suppressed by
ambipolar diffusion and a laminar wind solution is established
instead. Dashed vertical lines indicate the base of the wind, which we
define according to \citet{1993ApJ...410..218W} as the vertical
position, $z=z_{\rm b}$, where the rotation becomes
super-Keplerian. Since we define our azimuthal velocity, $v_\phi$ as
the deviation from the Keplerian background flow, this can be verified
in Fig.~\ref{fig:prof_OA-b5} by the fact that the Reynolds stress,
$T_{R\phi}^{\rm Reyn}\equiv \rho\,v_R\,v_\phi$, vanishes at this
point. The same is true for the vertical tensor component,
$T_{z\phi}^{\rm Reyn}$, which makes it convenient to define the
\emph{wind stress},
\begin{equation}
  T_{z\phi} \equiv 
   \left. T_{z\phi}^{\rm Maxw}\, \right|_{z=+z_{\rm b}}
  -\left. T_{z\phi}^{\rm Maxw}\, \right|_{z=-z_{\rm b}}
  \label{eq:wind_stress}
\end{equation}
at this point. For the $R\phi$ component of the stress tensor,
responsible for redistributing angular momentum radially, we
furthermore obtain average values within the disk body $z\in[-z_{\rm
    b},+z_{\rm b}]$,
\begin{equation}
  T_{R\phi} \equiv \frac{1}{2 z_{\rm b}}\,\int_{-z_{\rm b}}^{+z_{\rm b}}
  \left( T_{R\phi}^{\rm Maxw} + T_{R\phi}^{\rm Reyn} \right)\,{\rm d}z\,,
  \label{eq:accr_stress}
\end{equation}
which we list separately in Table~\ref{tab:results}, where values have
been normalized to the midplane pressure, $p_0$. Already for a
midplane plasma parameter as low as $10^5$, the wind stress exceeds
the $R\phi$ (accretion) stress by a factor of four. \BS also report
that the wind stress exceeds the accretion stress; for their fiducial
run, the discrepancy is bigger than an order of magnitude.

In the lower panel of Figure~\ref{fig:prof_OA-b5}, we see that in the
absence of MRI turbulence, the hydrostatic balance extends further up
in the disk and the gas pressure remains the dominant stabilizing
force all the way up to the base of the wind. Even in the magnetically
dominated wind layer, magnetic pressure gradients remain moderate and
the scale height of the disk remains roughly constant up to $z_{\rm b}$.

\begin{figure}
  \includegraphics[width=\columnwidth]{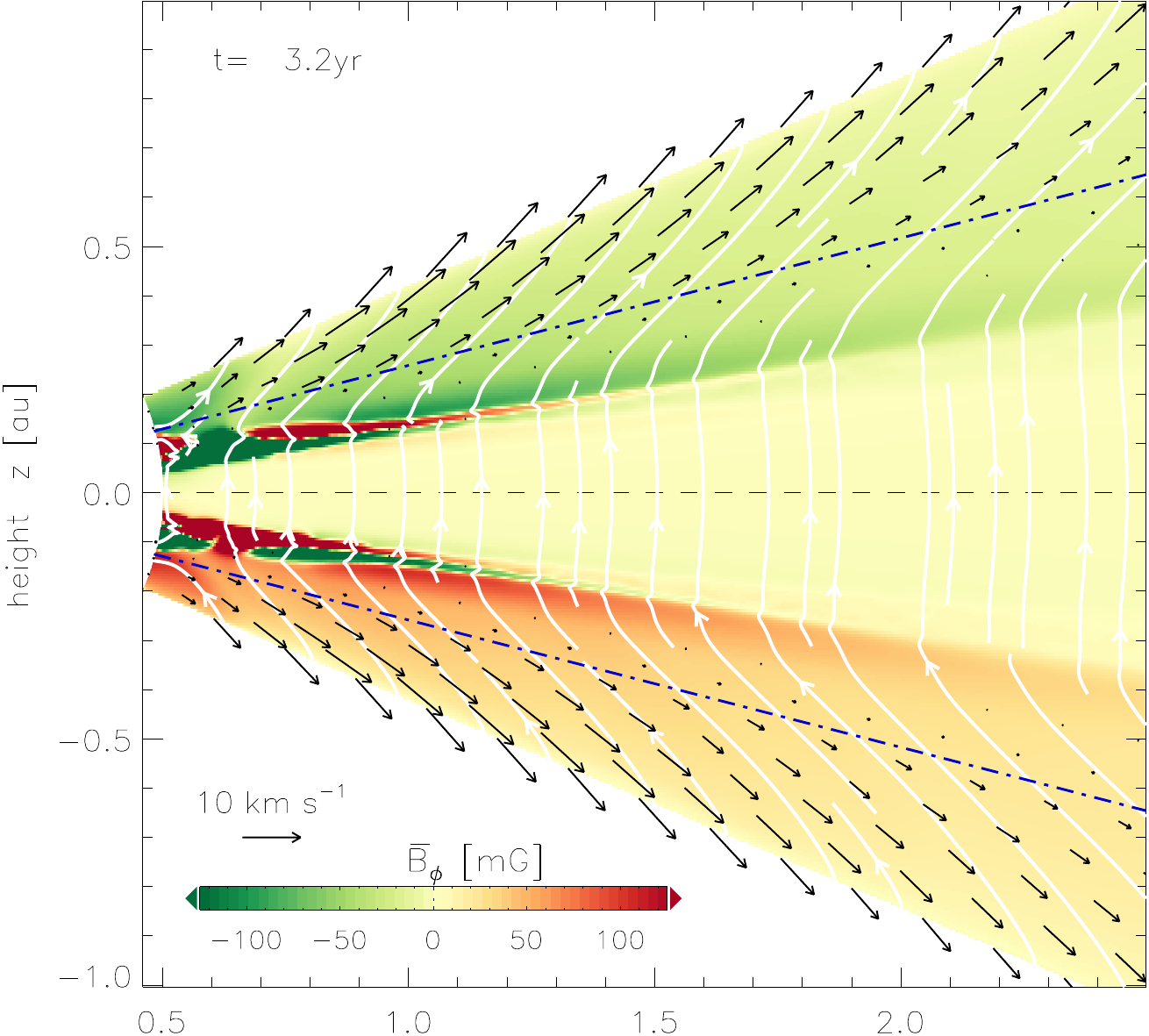}\\[6pt]
  \includegraphics[width=\columnwidth]{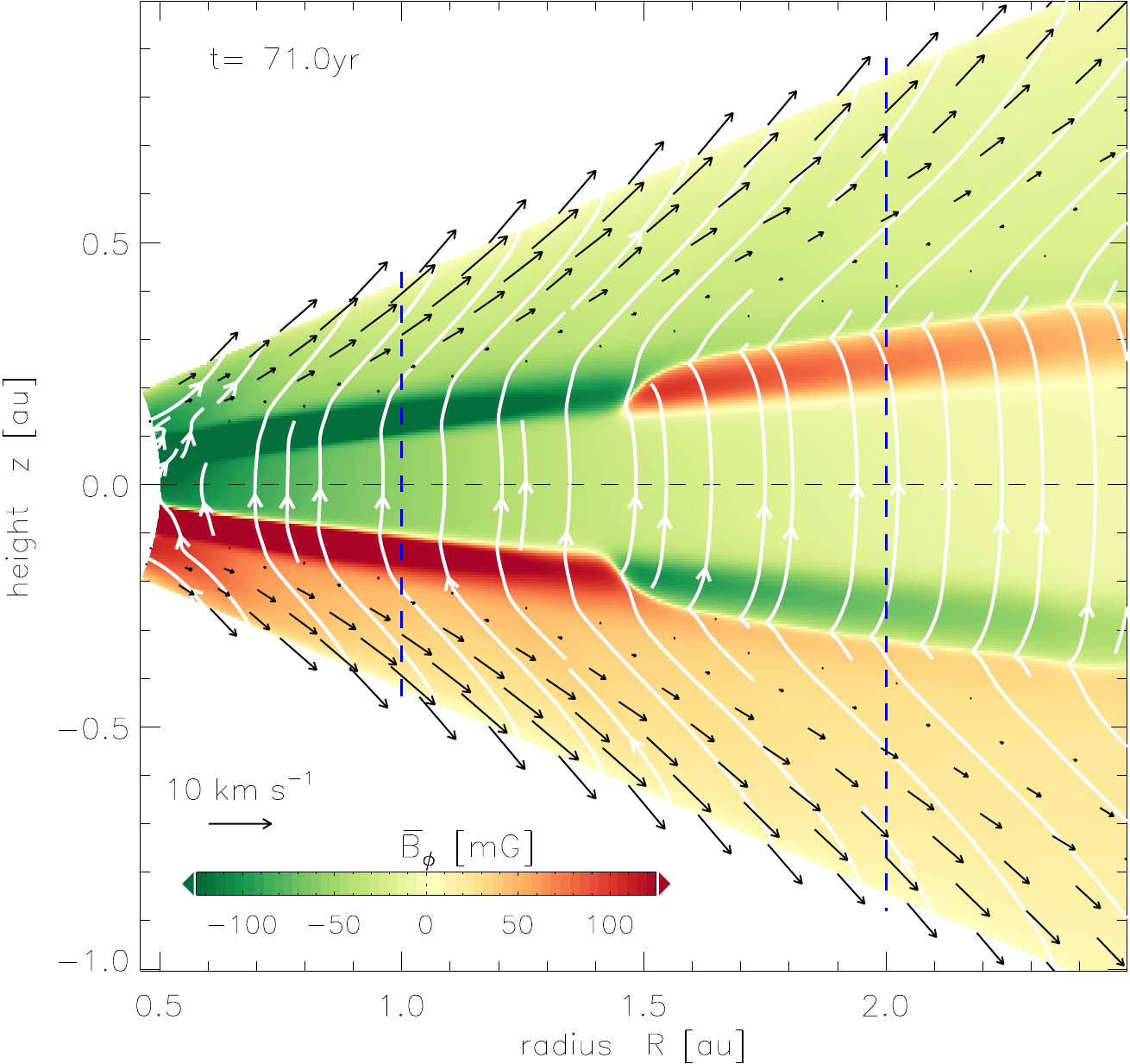}
  \caption{Field topology of our fiducial simulation at different
    evolution times. The azimuthal magnetic field (color) has been
    restricted to values $|B_\phi|< 125\mG$ for clarity; peak values
    are a few hundred ${\rm mG}$. We also show projected magnetic
    field lines (white) and velocity vectors (black). Additional lines
    indicate the position, $z_{\rm b}$, of the wind base (dot-dash),
    and the radial location of the profiles plotted in
    Figs.~\ref{fig:wind_OA-b5_r1} and \ref{fig:wind_OA-b5_r2} (dashed
    lines).}
  \label{fig:vis2D_OA-b5}
\end{figure}

In Fig.~\ref{fig:vis2D_OA-b5}, we show a zoom-in of the inner part of
the global field topology in our fiducial simulation OA-b5. The upper
panel shows the resulting field configuration early on, when
resistively modified MRI eigenmodes appearing inside $r=1.5\au$ are
still clearly discernible. Localized channel
solutions form at $z\approx 3\,H$ near the interface between the
magnetically decoupled region, characterized by $\Lambda_{\rm O} \ll
1$ and purely vertical field lines with $B_\phi\approx 0$, and the
AD-dominated intermediate layer. In this small region, both Elsasser
numbers are above unity (cf. Fig.~\ref{fig:elsa}), allowing for linear
growth of MRI channel modes. This transition layer is characterized by
a zigzag-shaped abrupt change in the field lines (see
Fig.~\ref{fig:mri_mode} below), associated with strong current
sheets. At this point, the laminar wind solution is already well
established in the FUV ionized surface layer, and has spread
throughout the radial extent of the simulation domain.

While the disk wind itself has already reached a steady state at this
point in time, the strong field layers continue to evolve. In the
lower panel of Fig.~\ref{fig:vis2D_OA-b5}, we accordingly show a later
evolution stage, where the relic MRI channels have morphed into strong
azimuthal field belts. These field belts can be understood as ``undead
zones'', that is, a region that does not sustain MRI, but where
diffusion can be balanced by the winding-up of preexisting radial
field via differential rotation \citep{2008ApJ...679L.131T}.  It
appears that the initial amplification of the magnetic field through
the growth of channel modes causes the ambipolar diffusion coefficient
to increase until it quenches further development of the MRI. This
notion is supported by a reference simulation, where the diffusion
coefficients $\eta_{\rm O}$ and $\eta_{\rm AD}$ were held \emph{fixed}
at their initial value, and where the MRI continues to produce
quasi-turbulent fields akin to the ones seen in
Fig.~\ref{fig:vis2D_O-b6} for the Ohmic-only case. The reversal of the
field direction seen at $r \simeq 1.5\au$ is a relic of the local
nonlinear development of the MRI modes early on (as may be seen in the
upper panel at $r\simeq 0.75\au$). This interface between the
azimuthal field belts of opposite parity moves radially outward at a
speed of $\approx 0.01\au\yr^{-1}$.

% ---

\begin{table*}\begin{center}
\caption{Summary of simulation results.
  \label{tab:results}}
\begin{tabular}{p{1.75cm}cccccccc}\hline
  & $z_{\rm b}$             & $z_{\rm A}$
  & $T_{R\phi}^{\rm Reyn}$  & $T_{R\phi}^{\rm Maxw}$  & $T_{z\phi}^{\rm Maxw}$
  & $\dot{M}_{\rm wind}$    & $\dot{M}_{\rm accr}$ \\[2pt]
  & $[H]$                   & $[H]$
  & $[10^{-6}p_0]$          & $[10^{-5}p_0]$          & $[10^{-5}p_0]$
  & $[10^{-8}\!\Msun\!\yr^{-1}]$ & $[10^{-8}\!\Msun\!\yr^{-1}]$
\\[2pt]\hline\\[-6pt]
 O-b6      & ---            & $7.60\pm 0.45$
           & $6.87\pm 14.4$ & $7.44\pm 0.95$ & ---
           & $1.58\pm 0.59$ & $0.04\pm 1.91$ \\[4pt]
 OA-b5     & $5.23\pm 0.07$ & $7.31\pm 0.17$
           & $3.63\pm 0.19$ & $2.22\pm 0.06$ & $9.82\pm 0.08$
           & $0.79\pm 0.01$ & $0.43\pm 0.01$ \\
 OA-b6     & $7.22\pm 0.48$ & $6.94\pm 0.37$
           & $\!\!\!-0.21\pm  0.18$& $0.79\pm 0.06$ & $0.58\pm 0.06$
           & $0.19\pm 0.06$ & $0.09\pm 0.02$ \\
 OA-b7     & $7.31\pm 0.70$ & $6.39\pm 0.16$
           & $0.07\pm 0.13$ & $<0.01$ & $0.22\pm 0.01$
           & $0.03\pm 0.01$ & $0.00\pm 0.03$ \\[4pt]
 OA-b5-d4  & $5.27\pm 0.07$ & $7.33\pm 0.18$
           & $0.11\pm 0.30$ & $2.88\pm 0.11$ & $9.88\pm 0.12$
           & $0.76\pm 0.02$ & $0.34\pm 0.04$ \\
 OA-b5-flr & $4.81\pm 0.03$ & $6.90\pm 0.31$
           & $0.26\pm 0.21$ & $1.78\pm 0.02$ & $14.3\pm 0.02$
           & $1.57\pm 0.01$ & $0.64\pm 0.02$ \\
 OA-b5-flr-nx & $4.78\pm 0.03$ & $7.50\pm 0.30$
              & $2.28\pm 9.24$ & $1.87\pm 0.04$ & $13.0\pm 0.04$
              & $1.58\pm 0.01$ & $0.64\pm 0.03$ \\
 OA-b5-nx  & $5.10\pm 0.04$ & $7.34\pm 0.13$
           & $0.94\pm 9.29$ & $1.89\pm 0.06$ & $7.84\pm 0.02$
           & $0.67\pm 0.01$ & $0.30\pm 0.04$ \\
 \hline
\end{tabular}
\end{center}
  \parbox[t]{2\columnwidth}{\footnotesize The vertical position of the
    base of the wind, $z_{\rm b}$, and the Alfv{\'e}n point, $z_{\rm
      A}$, are found independent on the radial location when measured
    in local scale heights, $H$. The viscous accretion stresses
    $T_{R\phi}$ are vertically integrated within $|z|\le z_{\rm b}$ --
    note the different units for Reynolds and Maxwell stresses. The
    wind stress, $T_{z\phi}$, is measured at $z=\pm z_{\rm b}$. All
    stresses depend weakly on radius; listed values are at $r=3\au$.}\medskip
\end{table*}

% ---

The emerging current layers are directly related to
resistively-modified vertically-global MRI channel modes -- see
\citet{2010MNRAS.406..848L}, and Fig.~\ref{fig:mri_mode} in
Section~\ref{sec:mri_current}.  With MRI channels potentially spanning
large radial extents, as in the Ohmic run shown in
Fig.~\ref{fig:vis2D_O-b6}, such layers may be fed diffusively with
magnetic field from the MRI-active inner disk. This is supported by
our present models, where current layers appear to emerge from inner
disk regions.  What determines the exact shape of the surviving field
configuration at late times is presently unclear. We speculate that
the particular realization seen in the lower panel of
Fig.~\ref{fig:vis2D_OA-b5} is not necessarily a unique solution given
the specific disk geometry and ionization model, but may to some
extend depend on the early evolution history.  However, simulations
that were initiated with different random seeds, but were otherwise
identical, only showed minor variations in the final field appearance.
The configurations observed at late times are moreover found to remain
quasi-stationary, at least during the evolution time of $100\yr$ that
we have investigated.

\subsection{Typical wind solutions} % ---

Perhaps the most noteworthy result from run OA-b5 is the spontaneous
emergence of the expected ``hourglass'' geometry for the magnetic field,
allowing magnetic tension forces to accelerate the wind.  For the
adopted vertical field polarity, the field must bend radially outward
near the upper and lower disk surfaces, and the azimuthal field should
point in the positive direction in the lower hemisphere and reverse
direction above the midplane, as observed in the upper panel of
Figure~\ref{fig:vis2D_OA-b5}.

\subsubsection{Robustness of the emerging wind geometry}

The shearing box simulations of \BS have reflectional symmetry in the
radial direction, that is, with respect to mirroring about
  $x=0$, and hence contain no information about the location
of the star. Instead of giving rise to a physical wind solution, their
fiducial run produces a radial field that possesses odd-symmetry about
the midplane, with an inward pointing field in one hemisphere and an
outward pointing one in the other. To test the robustness of the
emerging solution in our setup, we initiate several instances of run
OA-b5, using different random number seeds when perturbing the initial
velocity field, and in each case we recover the correct wind solution
with outward bending field lines. Our initial model has a net-vertical
field with a radial dependence such that the midplane value of
$\beta_{\rm p}$ remains constant. Because the gas pressure itself
decreases with radius, the corresponding $B_z(\Rc)$ (with
$\partial_{\Rc} B_z<0$) results in an \emph{outward} magnetic pressure
force, which moreover has a stronger acceleration effect in the
low-density upper disk layers. While a vertical flux distribution that
has the flux centrally concentrated can be expected when accounting
for inward advection and outward diffusion of flux
\citep{2014ApJ...785..127O}, our particular choice is of course
arbitrary. Radial magnetic gradients are furthermore known to affect
the outflow's mass loading in axisymmetric calculations where the disk
is a boundary condition \citep{2006MNRAS.365.1131P}.

In our simulations, the pressure gradient related to $B_z(\Rc)$ may
potentially affect the direction toward which the field lines first
bend from their starting configuration. We have tested this by
running reference models with different magnetic pressure
gradients. If $\partial_{\Rc} B_z>0$, the initial condition has an
unbalanced inward pressure force, and we do indeed find the field
lines to spontaneously bend \emph{inward}; this happens simultaneously
in the upper and lower hemisphere of the disk, such that the overall
symmetry that is obtained is again even. The launching of the inward
wind is restricted to the inner disk, probably because the vertical
field lines become too stiff (owing to $B_z$ increasing with radius)
to be suitable for wind launching in the outer disk.

For a neutral situation with $\partial_{\Rc} B_z=0$, we still observe
\emph{outward} bending of the field lines. Starting from the inner
radial domain, the outward propagating establishment of the wind
region is found to stall at some radius, whereas the wind was quickly
established throughout the entire domain in model OA-b5. As before
(for the case of an outward magnetic pressure gradient), this is
probably related to the field lines becoming too strong to support the
wind launching mechanism at large radii. The wind indeed stalls
further out in a run with weaker overall net-vertical field. This poses
the question whether the vertical profiles of $\Lambda_{\rm O}(z)$,
and $\Lambda_{\rm AD}(z)$ that we obtain from our ionization model put
strong constraints on permissible vertical field amplitudes as a
requirement for wind launching.

A possible reason for the outward bending, even in the case of neutral
magnetic pressure forces, may be that the vertical shear present in
the global models (because of the baroclinic conditions) naturally
bends the azimuthal component of the field in the correct direction
required by the physical wind solution. We have checked that the
outward bending of the vertical field lines is however equally seen in
(strongly flaring) \emph{globally} isothermal models that do not have
any vertical gradients in the rotation velocity or any radial
gradients in the vertical magnetic field. The outward pointing
configuration is energetically favored because setting it up involves
diluting rather than concentrating magnetic flux.

Our simulations demonstrate in any case that global models
spontaneously develop the correct field geometry, although we caution
that this conclusion needs to be tested in future simulations that
moreover adopt improved boundary conditions for the magnetic field at
the upper and lower (and potentially the inner radial) surfaces of the
simulation domain. Ultimately, the emergence of the wind geometry will
have to be studied in simulations that do not start from
well-motivated (but nevertheless arbitrary) initial conditions, but do
account for the formation of the PPD from a collapsing molecular cloud
\citep{2014arXiv1401.2219L}.

Lastly, for the parameters that we have considered here, we do not
find a self-limiting of the wind via shielding of FUV photons
\citep{2012ApJ...758..100B}.  This is consistent with the T~Tauri
scenario of \citet{2012A&A...538A...2P}, who performed disk wind
chemical modeling for various protostellar evolution stages. The
authors find the shielding of FUV photons to be important for the
Class~I and Class~0 cases.  Their T~Tauri case, with mass flow rates
comparable to our fiducial model, is sufficiently FUV-ionized to
reduce ambipolar diffusion enough so that the neutrals are swept up in
the wind out to at least a radius of $9\au$.

\begin{figure}
  \center\includegraphics[height=1.3\columnwidth]{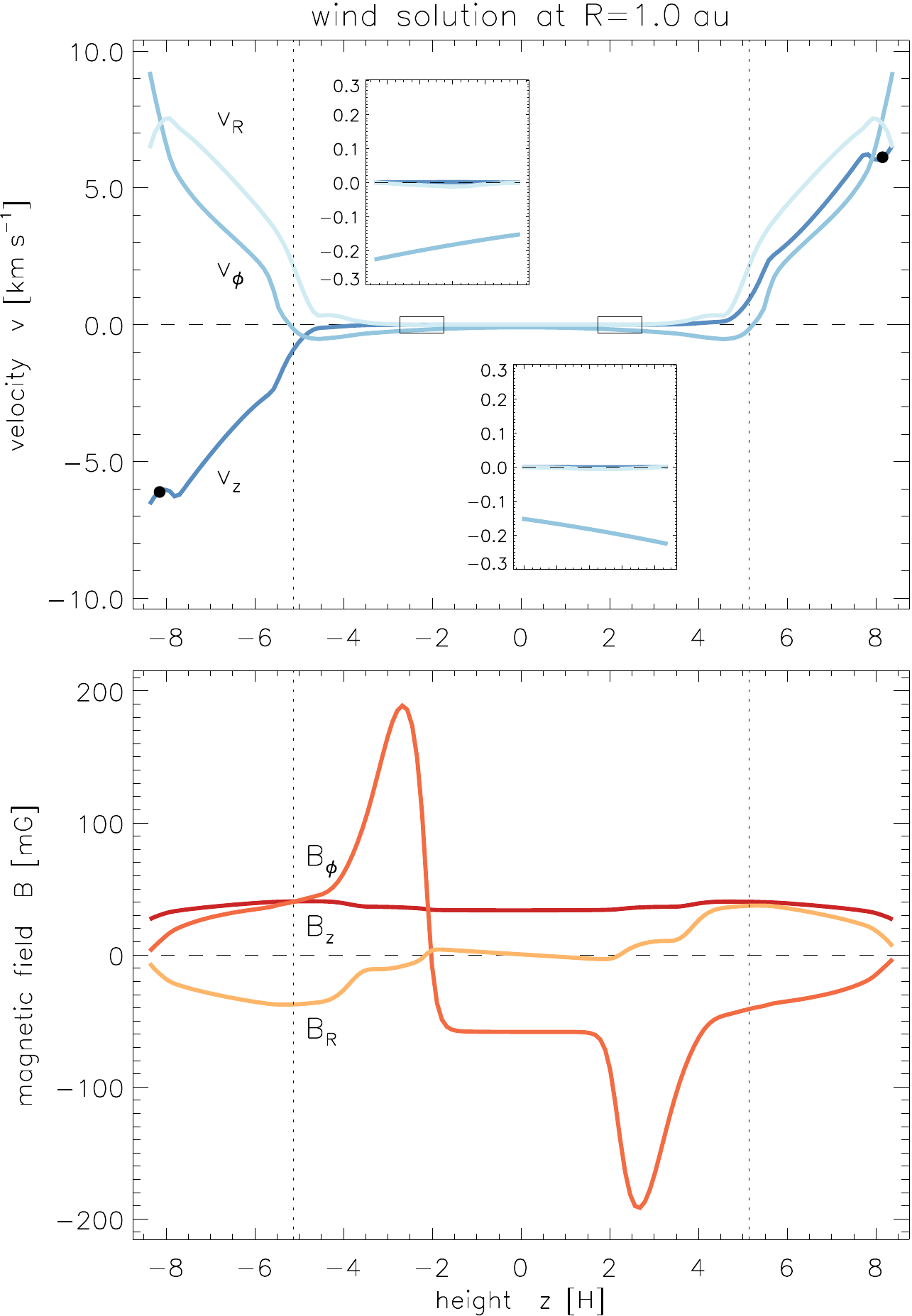}
  \caption{Vertical profiles of the velocity components (upper panel)
    and the magnetic field components (lower panel) projected to
    cylindrical components for run OA-b5 at spherical radius $r=1\au$
    (see the inner dashed line in Fig.~\ref{fig:vis2D_OA-b5}b). The
    Alfv{\'e}nic point with respect to $v_{\rm A\,z}\equiv
    B_z/\!\sqrt{\rho}$ is marked by filled circles. Dotted vertical
    lines indicate the base of the wind, defined as the point where
    $v_\phi$ becomes super-Keplerian \citep{1993ApJ...410..218W}.}\medskip
  \label{fig:wind_OA-b5_r1}
\end{figure}

\subsubsection{Comparison with local models}

Except for the reversal of the horizontal magnetic field components
(which at this stage appears to occur rather gently instead of within
a narrow current layer), the field structure observed in the upper
panel of Fig.~\ref{fig:vis2D_OA-b5} is similar to that described by
\BS and shown in their figure 10.
Near the midplane, where the dominance of the Ohmic resistivity
retards the growth of currents, the field is dominated by the vertical
component. Moving to higher altitudes, where the magnetic coupling
increases, the field bends outwards because a radial velocity is
generated through the azimuthal force balance between the Coriolis
force and magnetic tension. We enter the FUV layer at disk altitudes
$|z| \gtrsim 4.5 H$, coinciding with the base of the wind and the wind
itself, as illustrated by the velocity vectors.

\begin{figure}
  \center\includegraphics[height=1.3\columnwidth]{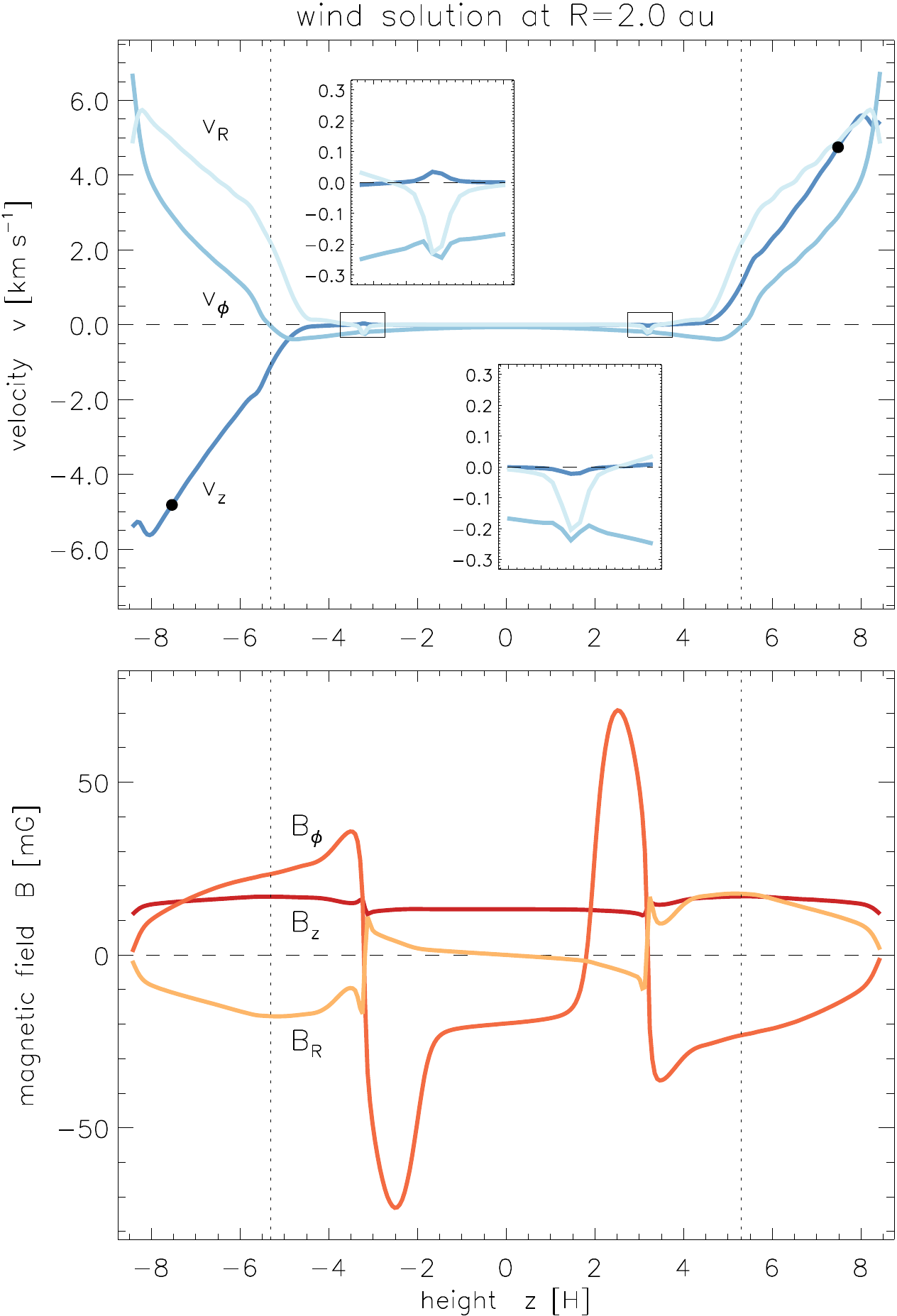}
  \caption{Same as Fig.~\ref{fig:wind_OA-b5_r1} but for $r=2\au$,
    where the horizontal field belts have opposite parity compared to
    the field within the wind layer. Inlays magnify the velocity
    profiles near the strong current layers, which reach a fair
    fraction of the local isothermal sound speed, $\cS=1.1\kms$.}
  \label{fig:wind_OA-b5_r2}
\end{figure}

At later times, the lower panel of Fig.~\ref{fig:vis2D_OA-b5} shows
that the varying polarity of the strong azimuthal field belts gives
rise to two distinct field line geometries. In the inner region
($r\simlt 1.5\au$) the orientation of the field belts is the same as
the one in the wind, and the vertical field lines bend smoothly at the
transition into the wind base. Because some negative $B_\phi$ has
diffused into the Ohmic-dominated midplane, there is a weak current
layer forming at $z\simeq-2.15\,H$. In contrast, further out at
$r\simgt 1.5\au$, the horizontal-field belts are anti-aligned with the
magnetic field direction of the wind. This can also be seen in the
curvature of the field lines, which are concave (towards the star) in
this region. To connect the different layers, the azimuthal and radial
field components have to change their direction within two thin
current layers located at $z\simeq \pm3.2\,H$. To study these
configurations in more detail, we plot separate vertical profiles for
$r=1\au$, and $r=2\au$ in Figs.~\ref{fig:wind_OA-b5_r1} and
\ref{fig:wind_OA-b5_r2}, respectively. For easy reference, the radial
positions are furthermore indicated in Fig.~\ref{fig:vis2D_OA-b5} by
means of dashed lines. The isothermal sound speeds, for the initial
disk model, are $\cS=1.5\kms$, and $1.1\kms$ at the two respective
radii.

The wind profile in the inner disk (see Fig.~\ref{fig:wind_OA-b5_r1}),
is qualitatively similar to the wind solutions previously found in the
quasi-1D simulations with ``even-z'' symmetry, plotted in fig. 10 of
\BS. The current layer is much weaker in our case, and the dip in
$v_r$ is barely visible in the left inset of
Fig.~\ref{fig:wind_OA-b5_r1}. This may be related to differences in
the ionization model resulting in $\Lambda_{\rm AD}>1$ in the region
around $|z| \simeq 3\,H$, in our disk model. Differences in the
particular diffusivity profiles also explain the pronounced peaks in
$B_\phi$ in this ``undead'' transition layer, which are not seen in
the simulations of \BS.

The profile at $r=2\au$, with double field reversal in $B_\phi$, is
illustrated in Fig.~\ref{fig:wind_OA-b5_r2}. Because of the strong
current layers that are emerging in this case, this configuration
resembles more closely the situation seen in the local models (again
cf. fig. 10 in \BS). Unlike in their simulation, we however observe
\emph{two} current layers, one on each side of the disk. These arise
because the development of MRI channel modes, and their subsequent
evolution into ``undead'' azimuthal field belts with opposite polarity
to the background wind, forces the horizontal field to change
direction three times. Strong current layers occur at altitudes where
the field and gas couple more strongly, namely at $|z| \sim 3
H$. Although the occurrence of strong azimuthal field belts is not
precluded in local simulations (as their existence may depend on the
details of the ionization model), it seems likely that the alternating
field belts we observe will only be a robust feature of global
simulations. Apart from their multiplicity, the current layers appear
very similar to the ones presented in \BS. The radial velocity $v_r$
shows a dip, while $v_\phi$ is modulated as a sine (the slight offset
of $v_\phi$ in our simulation is due to the radial pressure
support). The characteristic modulation (reflecting the abrupt kink in
the field lines) is clearly seen in $B_r$ and $B_\phi$, which agree
very closely with the $B_x$, and $B_y$ obtained by \BS. The emergence
of such strong current layers naturally begs the question of their
stability. Before we however discuss this issue in
section~\ref{sec:current}, we first want to take a look at the current
sheets' likely origin as remnants of MRI eigenmodes.

\subsection{Relation between current layers and channel modes} % ---
\label{sec:mri_current}

\begin{figure}
  \center\includegraphics[width=0.85\columnwidth]{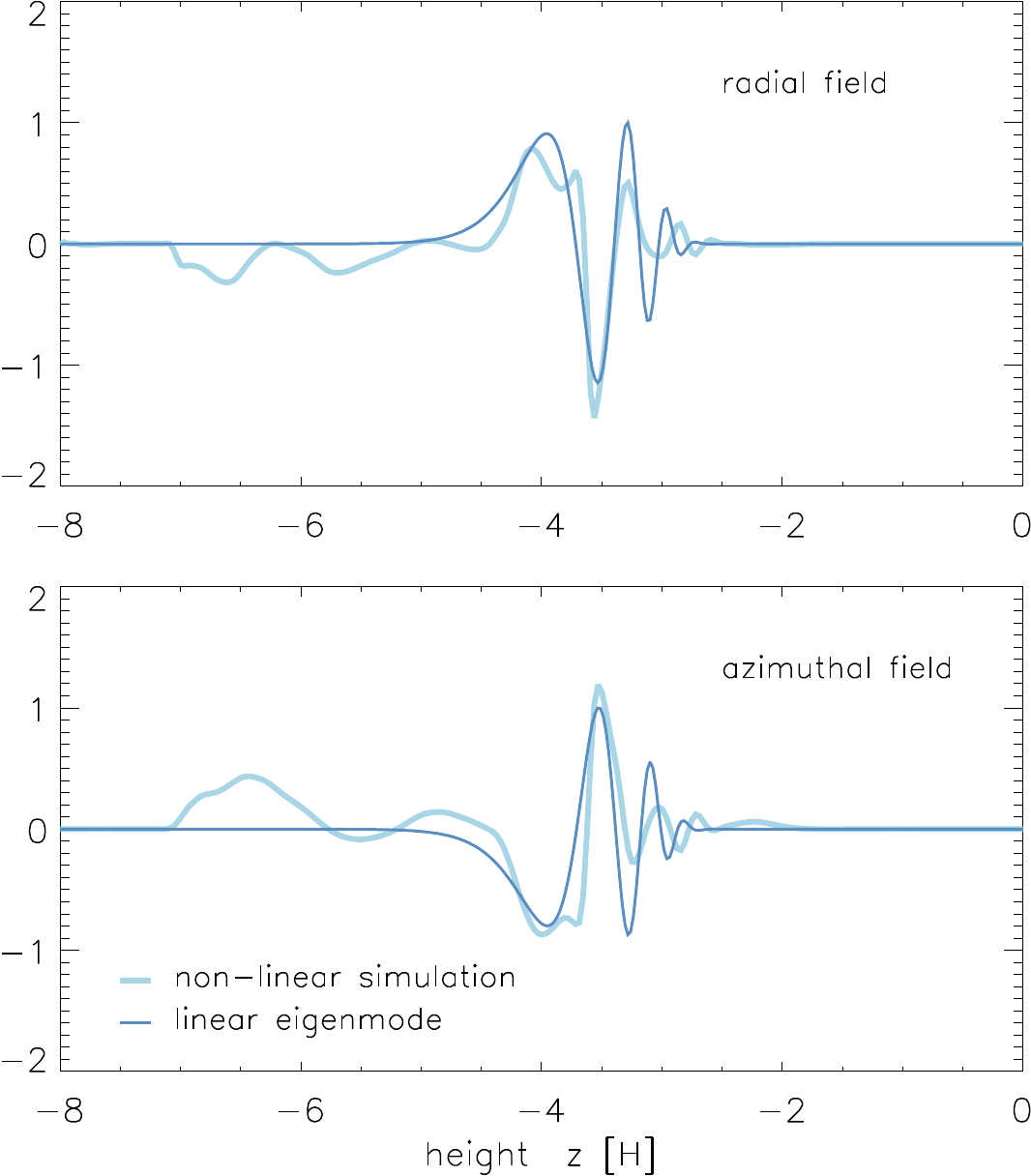}
  \caption{Vertical structure of diffusively modified MRI channel
    modes, with Ohmic resistivity only. The solution that has been
    extracted from the nonlinear global simulation (light color)
    agrees markedly well with the linear eigenmode solution (dark
    color).}\bigskip
  \label{fig:mri_mode}
\end{figure}

To draw a closer connection between localized MRI channel modes and
the current layers observed in our simulations, we compare the
structure of the fastest growing MRI eigenmode with the initial
evolution of our standard setup with $\beta_{\rm p\,0}=10^5$.  The
mode structure we find is qualitatively similar to that shown for a
spatially constant Ohmic resistivity \citep[cf. the appendix
  of][]{2013A&A...552A..71F}, and resembles the eigenmodes shown in
figure~4 of \citet{2014A&A...566A..56L}, who performed Hall-MHD
calculations that also included variable Ohmic resistivity.
Dissipation modifies the MRI eigenmodes
predominantly near the midplane, where the relative magnetization is
lowest and the MRI wavenumbers are the highest, explaining the
similarity of the solutions for constant $\eta$, and spatially varying
$\eta=\eta(z)$.

Here, instead of a stratified shearing box framework as previously
employed by \citet{2010MNRAS.406..848L}, we used the framework
introduced by \citet{VGSBpaper} to capture the full vertical structure
of the baroclinic initial condition.  The eigenmode analysis uses the
linear equations described in \citet[][Section~5]{VGSBpaper} with the
addition of terms for spatially varying Ohmic resistivity, and using
the functional forms and coefficients appropriate for the cylindrical
temperature structure of the disk initial condition
\citep[][Section~3.3]{VGSBpaper}.  The equations were discretized in
real space with Chebyshev cardinal functions on a Gauss-Lobatto grid
with 1000 points with the same vertical extent as the global
simulation ($\pm8$ scale heights).  Boundary conditions were enforced
by the ``boundary bordering'' method \citep{Boyd} forcing the magnetic
field perturbations to zero at the vertical edges of the domain.  The
analysis was performed at radial position $R=1\au$ where the Ohmic
resistivity profile, $\etao(z)$, used was extracted from the
simulation initial condition and interpolated to the grid used for the
eigenmode calculation.

We moreover obtained eigenmode solutions that additionally included a
fixed $\etad(z)$ profile, and these showed a nearly indistinguishable
mode structure. Because of the extra non-linearity introduced by the
$\beta_{\rm p}$ dependence of $\etad(z)$, we however restrict our
comparison to the Ohmic-only case.  In Figure~\ref{fig:mri_mode}, we
compare the early stage of our fully nonlinear global simulation with
the fastest growing eigenmode. The localized field reversals are also
clearly seen in the upper panel of Fig.~\ref{fig:vis2D_OA-b5}, which
shows the early evolution of model OA-b5 (however including AD).  To
isolate the linear MRI mode from the overlaid wind solution, we
subtract respective field profiles from two distinct snapshots during
the linear growth phase of the MRI.  This is possible because the wind
configuration emerges quickly, and then provides an approximately
stationary background in which the MRI grows. Differencing two
snapshots in time hence allows us to remove the wind part and extract
the exponentially-growing MRI eigenmode. While growing perturbations
near the midplane are high-wavenumber and thus efficiently damped, the
relative field strength in the disk halo is too strong to allow for
the MRI modes to fit within the available domain. This only leaves an
intermediate disk region near $|z|\simeq 3.5\,H$ for MRI channel
modes. In the Ohmic+AD simulations, this corresponds to the position
of the ``undead'' field belts (see lower panel of
Fig.~\ref{fig:vis2D_OA-b5}) with adjacent current layers. In those
simulations, the back-reaction of the growing eigenmodes on the
$\etad(z)$ profile makes the MRI self-quenching, leaving the current
layers as a remnant of the initial MRI channel mode.

\subsection{Stability of current layers and horizontal field belts} % ---
\label{sec:current}

\begin{figure}
  \begin{center}
    \includegraphics[width=0.95\columnwidth]{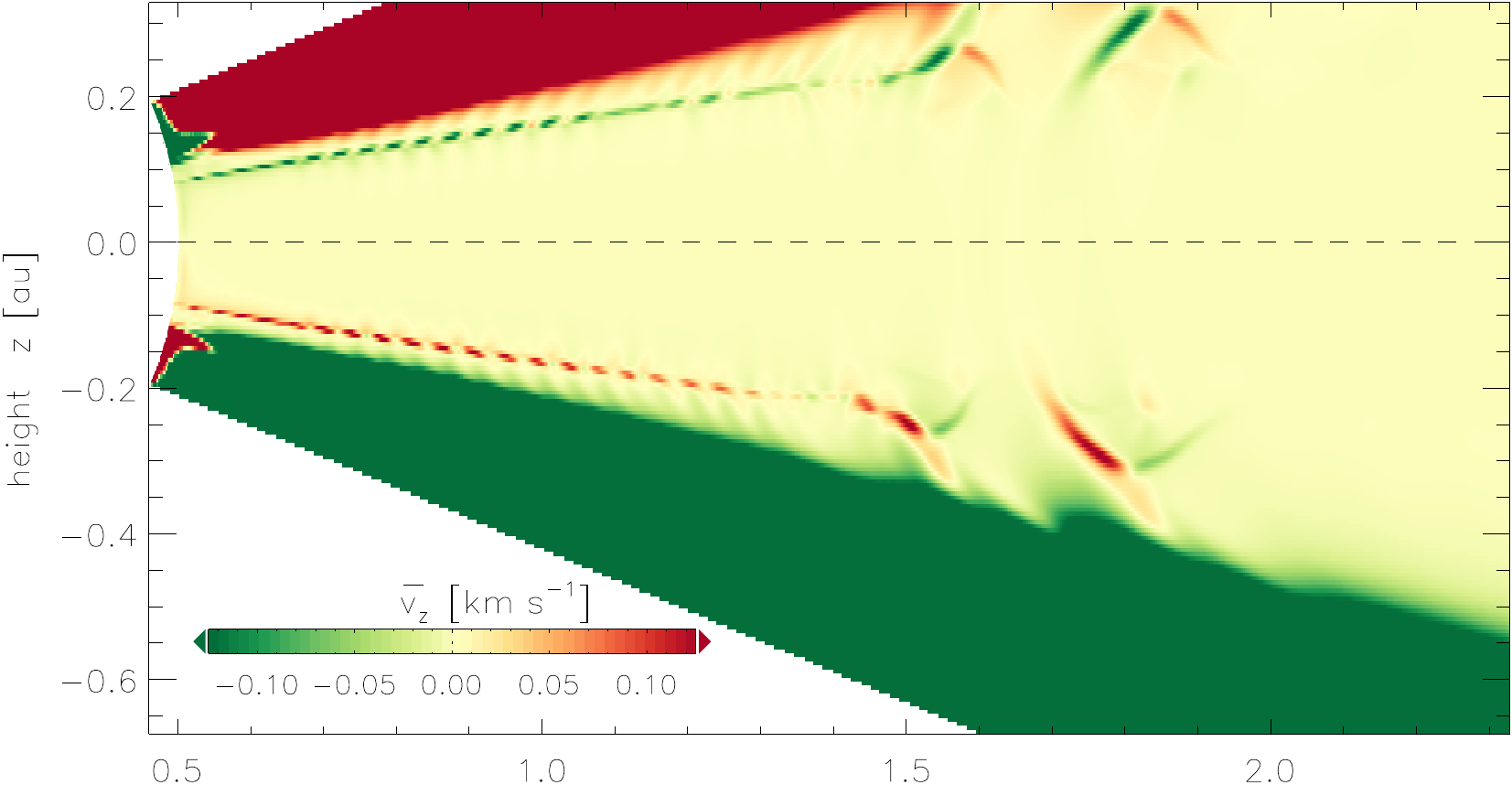}\\[4pt]
    \includegraphics[width=0.95\columnwidth]{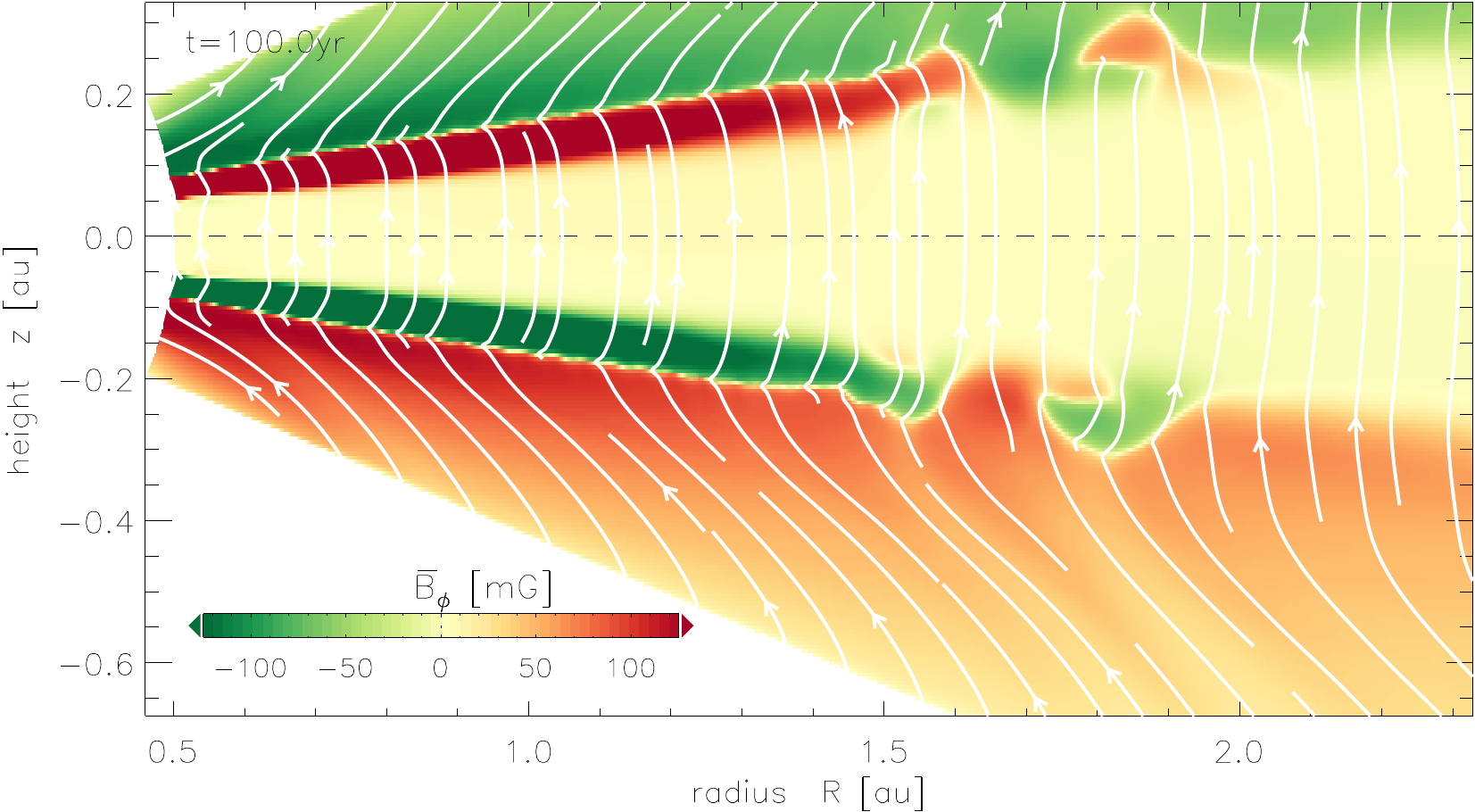}
  \end{center}
  \caption{Close-up showing the thin current layer of the flaring disk
    model OA-b5-flr breaking up into a KH-type
    instability. \emph{Top:} vertical velocity, $v_z$. \emph{Bottom:}
    azimuthal magnetic field strength, $B_\phi$, and poloidal magnetic
    field lines. The wavenumber of the perturbations is found to
    depend non-trivially on radius. Only for $r\simgt 1.5\au$, the
    nonlinear stage is clearly identified. The KH pattern is
    furthermore imprinted onto the mass-loading of the wind.}\medskip
  \label{fig:KH_OA-b5-flr}
\end{figure}

While we find the current layers to remain stable and long-lived in
model OA-b5, some of the other simulations which we will discuss below
also show signs of instability, leading to less regular fields.

\begin{figure*}
  \center\includegraphics[width=1.45\columnwidth]{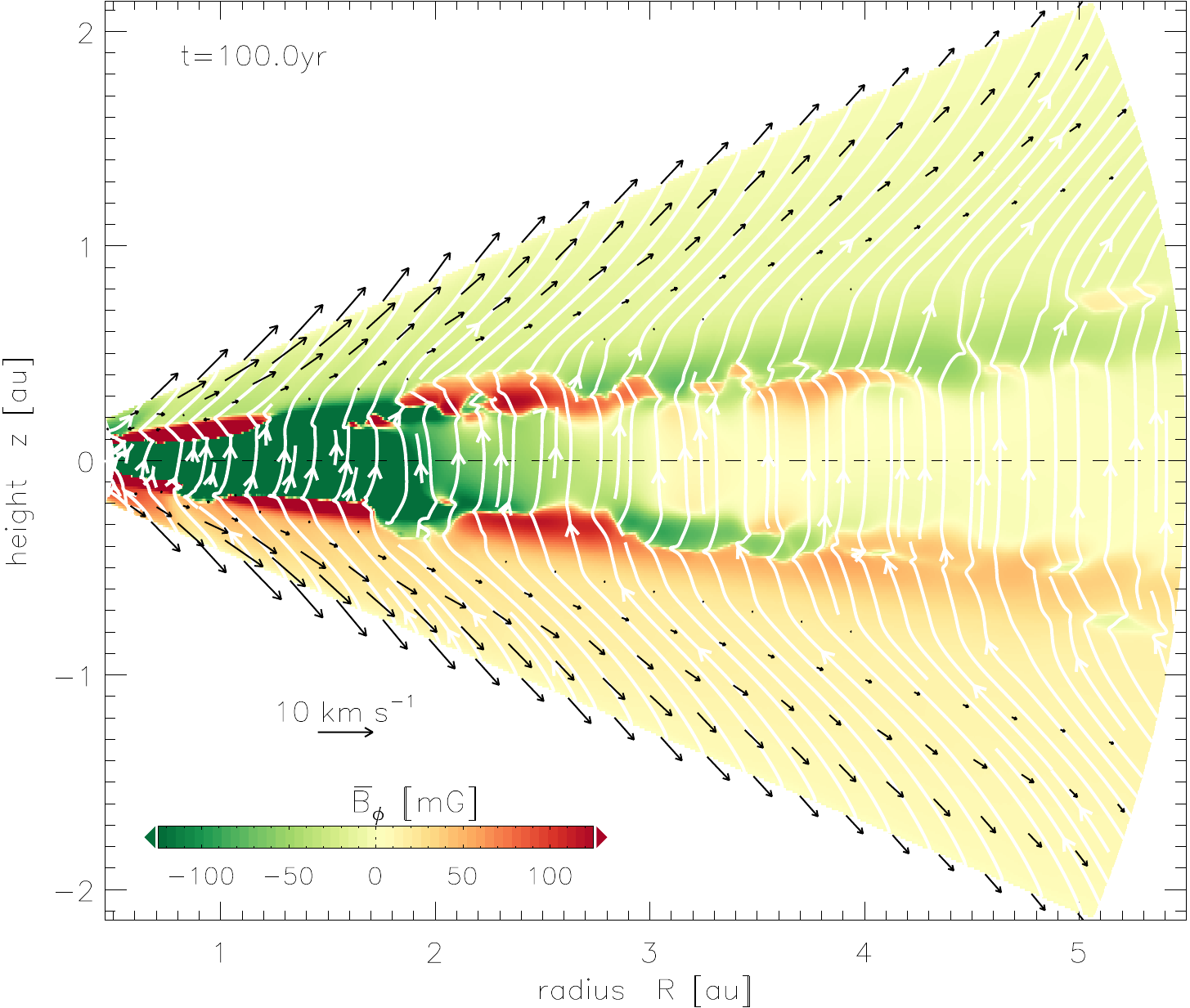}
  \caption{Same as figure Fig.~\ref{fig:vis2D_OA-b5}, but for model
    OA-b5-d4 with reduced dust fraction of $10^{-4}$, and showing the
    whole disk. Residual MRI fields form irregular patches below the
    wind layer leading to spatially intermittent current sheets.}\medskip
  \label{fig:vis2D_OA-b5-d4}
\end{figure*}

\subsubsection{Effect of disk flaring}

An example of this is illustrated in Fig.~\ref{fig:KH_OA-b5-flr},
where we show the vertical velocity, $v_z$ (upper panel), and the
azimuthal magnetic field, $B_\phi$ (lower panel), at the end of the
flaring disk simulation OA-b5-flr. Both color scales have been
restricted to highlight features in the current layer.  In this model,
the surface density interior to $1\au$ is smaller than in OA-b5, and
beyond $1\au$ it is larger, altering the ionization structure
sufficiently to put the current layer in a different regime. In
particular, the increased ionization in the inner disk apparently
allows the azimuthal field belts that form at intermediate heights to
reach larger amplitudes.  This, combined with the fact that these
field belts are of opposite polarity relative to the background wind
field, causes the two current layers located at $z\simeq \pm 3H$ to be
stronger and to support narrow, radially directed flows, or ``jets'',
similar to those in run OA-b5, as illustrated in
Figure~\ref{fig:wind_OA-b5_r2}.

Inside $r\simeq 1.4\au$, the velocity near the current layer shows a
distinct radial wavenumber, suggesting that the shear associated with
the radial jets leads to growth of a Kelvin-Helmholtz (KH)
instability. The pattern, however, remains more or less stationary,
indicating that the instability is not growing significantly into the
nonlinear stage. Although not obvious in Fig.\ref{fig:KH_OA-b5-flr},
wave-like perturbations to the magnetic field are also present in the
current layer. The possibility of current-driven tearing modes being
present also arises, but analysis of the shear rate associated with
the radial jets suggests that the observed instability is driven by KH
modes. The velocity associated with the shear is super-Alfv{\'e}nic,
and is therefore expected to stabilize the tearing modes that may in
principle grow in a shearing background \citep{1997JGR...102..151C}.

The wavelength of the velocity perturbations and magnetic field
changes abruptly for $r\simgt 1.4\au$, where now a characteristic
vortex-pair pattern emerges. The exact reason for this abrupt
transition is unclear, since most characteristic quantities, such as
the Elsasser numbers $\Lambda_{\rm O}$, and $\Lambda_{\rm AD}$, only
depend weakly and continuously on radius. The most likely explanation
may be the radial variation of the azimuthal field (inside the
``undead'' belts), which drops in strength around this radial
location. This is accompanied by a weakening and widening of the
current layer, and a significant reduction in the inflow velocity of
the radial accretion flow associated with the current layer, and a
corresponding reduction in the shear relative to the background
fluid. This suggests that the abrupt change in the nature of the
disturbance observed at radius $\sim 1.8\au$ is unlikely to be driven
by a KH instability, but the weakening of the shear may allow a
tearing-mode instability to develop instead. The appearance of
disturbances that have the character of growing magnetic islands
provides some support for this interpretation, but this is clearly a
tentative explanation that will require further investigation in
future work. The presence of a background wind, Keplerian shear,
ambipolar diffusion and a radially-directed inflow associated with the
current layer make an analysis of this problem somewhat complicated,
and beyond the scope of this paper. We note, however, that the local
studies on the development of parasitic modes in resistive MHD
\citep{2009MNRAS.394..715L,2010ApJ...716.1012P} may provide some
insight, even though ambipolar diffusion is neglected in those
studies.  \citet{2010ApJ...716.1012P} has analyzed MRI parasitic modes
in the presence of finite Ohmic resistivity and viscosity. In this
paper, $\Lambda_{\rm O}$, is identified as the relevant dimensionless
number (see their figure~11), with tearing modes being the dominant
parasitic modes for $\Lambda_{\rm O}\simlt 1$, and Kelvin-Helmholtz
modes dominating otherwise. Considering both $\Lambda_{\rm O}$, and
$\Lambda_{\rm AD}$, we assess that our simulations are situated near
this transition point.

\subsubsection{Effect of reduced dust fraction}

Figure~\ref{fig:vis2D_OA-b5-d4} shows the final appearance of the
entire disk for model OA-b5-d4, which has a reduced dust fraction of
$10^{-4}$ compared to a value of $10^{-3}$ in model OA-b5. On one
hand, as the reader may check from Table~\ref{tab:results}, the two
models are rather similar in their bulk properties, especially those
related to the wind.  This is unsurprising, since the reduced
cross-section for recombination on grains affects the ionization
fraction significantly only at large gas columns (see
Fig.~\ref{fig:elsa}), below the base of the wind.  The grains are less
important for the ionization state in the FUV layer, where the
electrons are so abundant that many remain free once the grains charge
to the Coulomb limit \citep{2009ApJ...698.1122O}.

In contrast to our fiducial model, the transition between the base of
the wind and the Ohmic-dominated disk interior is very chaotic in
model OA-b5-d4, whereas it is much more stable in model OA-b5. It
appears that the larger values of $\Lambda_{\rm O}$ and $\Lambda_{\rm
  AD}$ in this model at heights $z \sim \pm 2.5H$ allow the channel
modes to grow further into the nonlinear regime, such that parasitic
modes are able to develop and cause the field belts to break up into
regions with locally coherent field.  The field amplification during
this phase means ambipolar diffusion strengthens, as in run OA-b5.
Eventually the local evolution is driven by a combination of field
winding and diffusion.  The end result is the creation of local
``undead'' patches that evolve on timescales that are comparable to
the run duration.

The different evolution of the horizontal field belts observed in runs
OA-b5-flr and OA-b5-d4 compared with OA-b5 show that the evolution of
the disk interior depends critically on the precise disk model that
determines the Elsasser number profiles. In particular, the ionization
state in the region between the Ohmic-dominated disk midplane and the
AD-dominated intermediate disk layers depends sensitively on various
model parameters. This suggests that a careful parameter study is
required to obtain a reliable picture of how PPDs evolve over their
life-times -- especially since their properties change substantially
over these time scales.  Such a parameter study would appear to be
most warranted when additional physics is included in the model, such
as the Hall effect and more realistic thermodynamics. In a situation
where the gas is allowed to be heated by resistive effects, the
ionization balance may be shifted towards higher conductivity in the
vicinity of the current layers, potentially leading to intermittent
behavior \citep*{2014A&A...564A..22F,2014ApJ...791...62M}.

\subsection{High-wavenumber instability at FUV transition} % ---
\label{sec:corona}

\begin{figure}
  \center\includegraphics[width=\columnwidth]{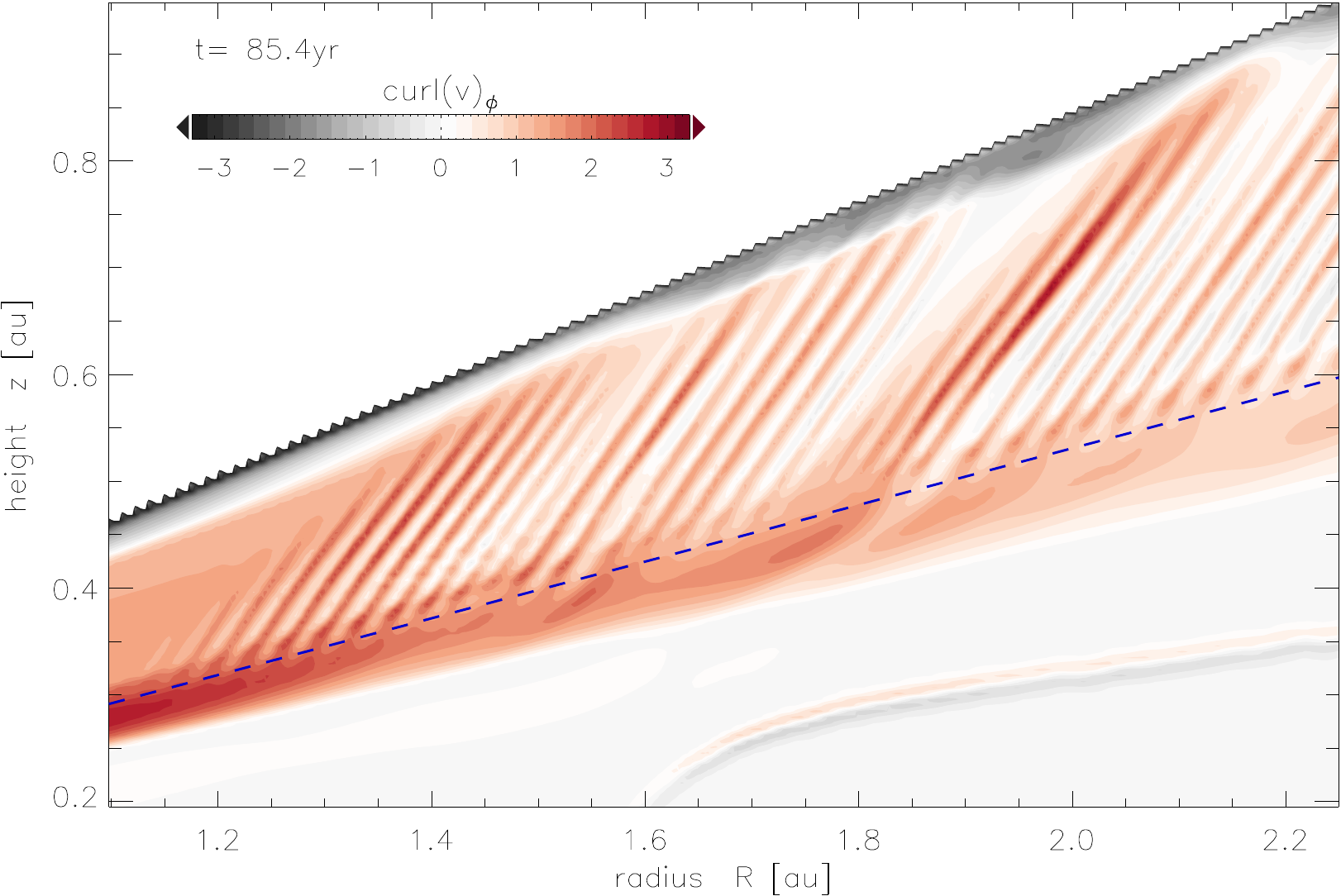}
  \caption{Color-coded azimuthal component of the flow vorticity (at
    late time in model OA-b5), tracing high-frequency perturbations in
    the magnetically dominated disk corona. Perturbations originate at
    the base of the FUV layer (dashed line), and are aligned with the
    field lines of the wind solution.}
  \label{fig:fishbone}
\end{figure}

A distinct instability emerges in the magnetically dominated corona of
the laminar disk (see Fig.~\ref{fig:fishbone}). The instability is
found to grow close to the smallest available wavenumber, and is only
present in the highly-resolved axisymmetric simulations. The
perturbations are roughly aligned with the outward-pointing field
lines of the laminar wind solution.  Similar patterns have been
observed in 2D-axisymmetric shearing box simulations with strong
vertical fields performed by \citet{2012A&A...548A..76M}, who also
finds field-aligned structures on the shortest length scales resolved
on the grid. The instability also bears resemblance with the ``striped
wind'' seen by \citet*{2013A&A...550A..61L}. Similar to the
instability seen by \citet{2012A&A...548A..76M}, we find a moderate
modulation of the vertical mass flux by the instability pattern. We
however do not find material falling down along the field lines as
reported there for certain parameters.

We identify three candidate mechanisms for creating such disturbances:
(i) the vertical-shear instability investigated recently by 
\citet*{2013MNRAS.435.2610N}, potentially modified by anisotropy from 
field-line tension, (ii) a Kelvin-Helmholtz type instability related 
to the vertical gradient of the horizontal velocity near the base of 
the wind, and (iii) a buoyant interchange instability of the azimuthal 
field. It is important to note that the KH-type instability can create 
the elongated aspect ratio seen in Fig.~\ref{fig:fishbone}, when 
perturbations are carried-away by the overlaid laminar wind velocity -- 
and we indeed find this to be the case in situations where the 
instability occurred before the wind was established. The conclusive 
hint that a Kelvin-Helmholtz instability causes the pattern in
Fig.~\ref{fig:fishbone} is a strong velocity gradient at the exact
location of the transition into the FUV layer. This gradient can be
clearly identified in Fig.~\ref{fig:wind_OA-b5_r2} (situated just
above the wind base at $z_{\rm b}$).  The shear layer manifests itself
as an additional steepening of the wind profiles, and can also be seen
in the corresponding plot of \BS (see their figure~10).

The physical origin of this sharp velocity gradient can be traced back
to a peak in the ambipolar part of the electromotive force in the
induction equation. Because the coefficient $\eta_{\rm AD}(r,\theta)$
sharply drops at the position of the FUV interface, the term $\nabla
\times (\etad \,[\,(\nabla\tms\B)\times\bb\,]\times\bb)$ can become
significant in the induction equation even for smoothly varying (or
constant) $\B$. In the steady state, this curl of the ambipolar
electromotive force needs to be balanced by the induction term,
$\nabla\times(\V\times\B)$, requiring as a consequence a significant
shear gradient in the horizontal velocity (again assuming a smooth
$\B$). Given its strong dynamical effect, this raises the question
whether the coincidence of the wind base and the FUV interface is in
fact accidental as claimed by \BS.

In our disk model, the FUV layer is imposed \emph{ad hoc}, with a
prescribed column density (Section \ref{sec:fuv}). The resulting
interface is smeared out to a prescribed width. Reflecting the
predominant \emph{direct} origin of the FUV photons, while scattering
is expected to remain negligible, the tapering is done in the radial
direction, and as a function of column density (rather than
position). As a consequence, the resulting $\eta_{\rm AD}(r,\theta)$
may develop sharp steps on the numerical grid in the
$\theta$~direction -- in turn resulting in sharp velocity
gradients. We accordingly ran a reference model, where the FUV
transition was softer. In this model, we did not observe a strong
velocity gradient near the wind base, and also found the instability
to be largely suppressed. Also the flaring-disk model, OA-b5-flr, did
not show any signs of the instability. This can potentially be
explained by the FUV transition following a \emph{concave} line in
this case, producing a different aliasing of the tapering function on
the spherical-polar mesh. We conclude that the instability seen in
Fig.~\ref{fig:fishbone} is likely caused by the particular numerical
parametrization of the FUV layer in our model, and is entirely
unrelated to the ``striping'' instability seen in
\citet{2012A&A...548A..76M} and \citet{2013A&A...550A..61L}.  The
instability may be be triggered by numerical effects at the grid
scale, in a manner analogous to how physical secondary
Kelvin-Helmholtz instabilities can be triggered by purely numerical
effects as a shear layer thins to the grid scale
\citep{2012ApJS..201...18M}.  We remark that in the case of this wind
instability too the mechanism behind the instability is genuinely
physical, and may occur in real PPDs where there is a sharp local
change in the ionization state.

\subsection{Non-axisymmetric simulations} % ---
\label{sec:3d}

\begin{figure}
  \begin{center}
    \includegraphics[width=0.8\columnwidth]{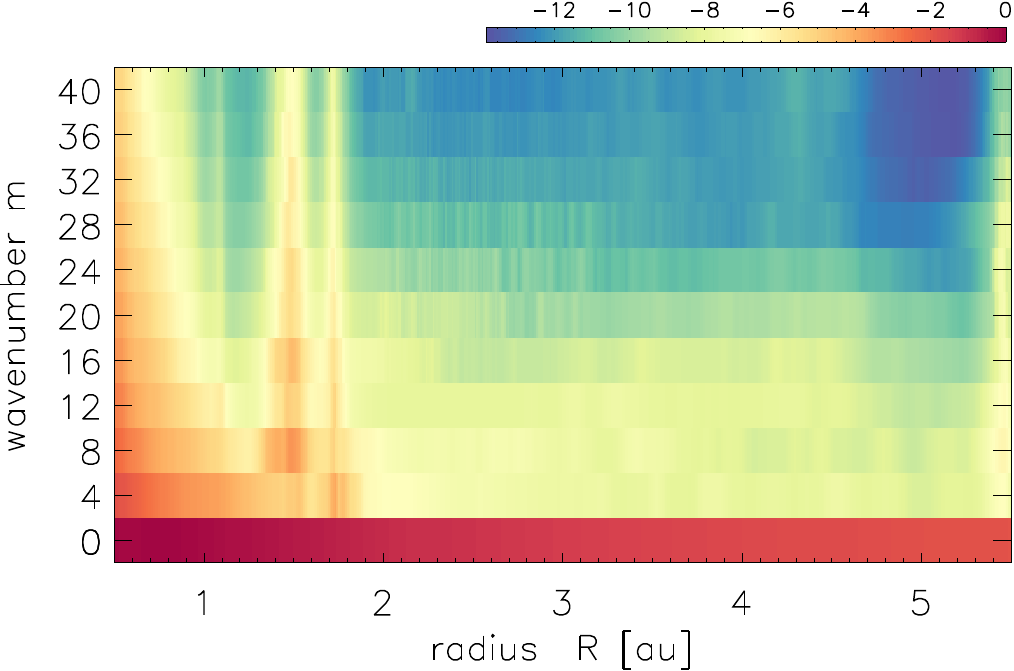}\\[10pt]
    \includegraphics[width=0.8\columnwidth]{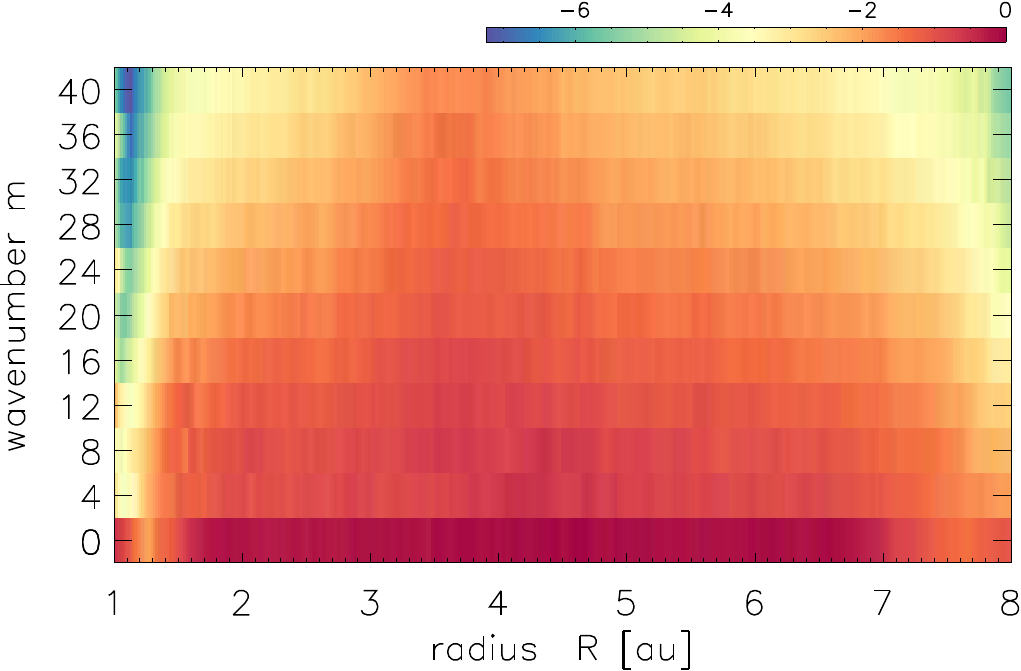}
  \end{center}
  \caption{Vertically-averaged azimuthal power spectra of the
    $B_\theta$ magnetic field. \emph{Top:} Model OA-b5-flr with a
    laminar wind. \emph{Bottom:} A run with MRI active surface layers
    in a domain of slightly different radial size, from
    \citet{2013ApJ...779...59G}. Owing to the restricted azimuthal
    extent in both runs, the only wavenumbers present are multiples of
    four.}
  \label{fig:mpower}
\end{figure}

The presence of secondary instability mandates investigating the
non-axisymmetric evolution. This is done exemplarily for the two
  models OA-b5, and OA-b5-flr, whose 3D counterparts OA-b5-nx, and
  OA-b5-flr-nx are also listed in Table~\ref{tab:results}, showing
  only minor deviations from their respective 2D runs. Given the
moderate magnetic Reynolds numbers in these regions, it is not
\emph{per se} obvious whether non-axisymmetric
instabilities can transfer significant amounts of energy into modes
with higher azimuthal wavenumber. To address this question, in
Fig.~\ref{fig:mpower}, we show vertically-averaged azimuthal power
spectra of the magnetic field. The color-coding indicates the
logarithm of the power spectrum of $B_\theta$ as function of
wavenumber, $m$, and radial position in the disk. In the upper panel,
we show the result for model OA-b5-flr. With the exception of the
Kelvin-Helmholtz features (between $R=1.5-2.0\au$) also seen in
Fig.~\ref{fig:KH_OA-b5-flr}, almost all power resides in $m=0$. Even
in the regions of the secondary instabilities, the power in modes with
$m>0$ remains very small. The power seen adjacent to the inner radial
domain boundary is caused by reducing the diffusion coefficients
there, which is an attempt to mimic the MRI-active inner disk (see
Section~\ref{sec:bcs} for details).

This can be contrasted to the case of the classic layered PPD with
MRI-active surface layers \citep[see model M1
  in][]{2013ApJ...779...59G}, which we plot in the lower panel of
Figure~\ref{fig:mpower}. Even with the laminar midplane layer, a
significant amount of power is found for high-$m$ modes (note the
different color bar in this figure). The lack of power near the radial
boundaries is owing to a dissipative buffer region.
\citet{2013ApJ...779...59G} focused on the effects of a planet, which
they inserted after the time shown here. We hence conclude that in the
simulations that are dominated by axisymmetric configurations of
AD-laminar winds, secondary instabilities do not lead to significant
power in non-axisymmetric perturbations. This may potentially change
in the presence of stronger current layers that are prone to nonlinear
stage of parasitic modes as, for instance, can be seen for model
OA-b5-d4 with a stronger dust depletion. When combined with stronger
net-vertical fields and flaring, these dust-depleted disks may develop
intermediate layers that are on the verge of becoming turbulent.

% ------------------------------------------------------------------------------

\section{Summary}
\label{sec:summary}

We have presented the first quasi-global MHD simulations of PPDs that
include Ohmic resistivity and ambipolar diffusion. In agreement with
the shearing box simulations of \BS, we find that the inclusion of
ambipolar diffusion has a dramatic effect on the evolution. Our main
results may be summarized as follows.

\begin{enumerate}[leftmargin=3ex]
\itemsep4pt

\item{In simulations where we include a weak vertical magnetic field
  and Ohmic resistivity, but neglect ambipolar diffusion, we obtain
  evolution that is consistent with the traditional dead zone picture:
  surface layers that sustain MRI-driven turbulence and a
  quasi-laminar midplane region.}

\item{The inclusion of ambipolar diffusion in models with weak
  magnetic fields, where the midplane value of $\beta_{\rm P} > 10^5$,
  results in quenching of the surface layer turbulence.  Such disks
  display very weak internal motions and negligible angular momentum
  transport.}

\item{Our most important result arises when we increase the strength
  of the magnetic field so that $\beta_{\rm P}=10^5$. This leads to a
  disk which is entirely laminar between 1--$5\au$, and, despite the
  still relatively weak field, to the launching of a
  magnetocentrifugal wind from the upper ionized layers, resulting in
  a significant wind stress being exerted on the disk that drives
  inward mass flow at a rate $\gtrsim 10^{-8} \Msun \yr^{-1}$.}

\item{We have tested the robustness of the wind solution by running
  multiple simulations with modified initial conditions. Because of
  the outward pressure force from the initial $B_z(\Rc)$, and the
  vertical differential rotation, we generally obtain magnetic fields
  that point outwards in the radial direction in both hemispheres, as
  required for a physical wind solution that extracts angular momentum
  from the disk. This is with the exception of a model with
  $\partial_{\Rc} B_z>0$, where an inward pointing wind was found.}

\item{The calculations treat the time-varying absorption of the
    FUV radiation from the star and its vicinity in the intervening
    material, yet there is no sign that absorption by the wind limits
    the ionization of the launching layer for mass loss rates in the T
    Tauri range.}

\item{Strong horizontal field belts are formed as remnants of
  stagnated MRI modes at intermediate heights in our disk models, and
  are likely maintained through a balance between field winding and
  ambipolar diffusion. These are found to have field polarities that
  may be aligned or anti-aligned with the background field associated
  with the wind, affecting the number of current layers that arise in
  the disk because of local field reversals. The belts polarity may
  depend sensitively on the early evolution of the MRI.}

\item{The current layers drive substantial accretion flows that are
  narrowly localized in height. The associated shear can be strong
  enough to allow small-scale Kelvin-Helmholtz instabilities to grow.}

\item{The global evolution of the horizontal field belts is found to
  depend on the Ohmic and ambipolar Elsasser numbers (and hence the
  ionization fraction). In our fiducial model, the belts alternate in
  polarity over large radial extents, but otherwise appear to be
  stable and only evolve on time scales that are long compared to the
  dynamical time scales.}

\item{Reducing the dust grain abundance by an order of magnitude, or
  allowing the disk to have a flaring structure, leads to more complex
  evolution in which the belts break-up into smaller-scale islands of
  locally coherent horizontal field polarity. We hypothesize that this
  evolution results from the increased ionization fraction that allows
  MRI channel modes to grow further into the nonlinear regime, where
  they are prone to break up \emph{via} parasitic modes.}
\end{enumerate}

Our global simulations are most comparable to the shearing box
simulations of \BS, and so it is worth briefly comparing our results
with theirs. Our results for weakly magnetized disks ($\beta_{\rm P} >
10^5$) with ambipolar diffusion and Ohmic resistivity compare well
with their results for disks with no net vertical magnetic field, that
is, no sustained wind and very weak turbulence and angular momentum
transport. This indicates that there is a clear lower limit for the
net vertical magnetic field required to drive an accretion flow that
is in agreement with observations (given our restricted assumptions
about the disk microphysics). Our results are in complete qualitative
agreement with \BS for higher magnetic field strength: disks are
laminar and a magnetocentrifugal wind is launched from the highly
ionized surface layers. The associated wind stress is sufficient to
drive an accretion flow that matches typical rates measured to be
reaching the surfaces of young stars.

Unlike earlier studies of magnetocentrifugal wind launching from
deeper towards the midplane \citep{1982MNRAS.199..883B,
  1983ApJ...274..677P, 1993ApJ...410..218W}, there is no requirement
for a strong magnetic field to be present in the disk when the wind is
launched from high altitudes, since the requirements for field
rigidity are met for globally weak fields in these low density
regions. The appearance of long-lived horizontal field belts is not
observed in the shearing box simulations of \BS, but this is because
of modest differences in the disk models and assumed ionization
physics (their model has a somewhat larger surface density at $1\au$,
and we include a direct component to the disk illumination by stellar
X-ray and FUV photons). The models agree qualitatively on the
appearance of strong current layers, but the stability of these could
not be examined by \BS as their models in which the field geometry
gave rise to these current layers used a quasi-1D approximation.  This
latter point illustrates the primary difference between the global and
local shearing box simulations. The global simulations appear to
produce and sustain the correct physical wind geometry spontaneously,
whereas the symmetries associated with the shearing box normally lead
to a radial magnetic field that points inwards in one hemisphere and
outwards in the other, such that the correct geometry needs to be
enforced. The potential role of the rather arbitrary initial
conditions, and the vertical boundaries in setting the magnetic field
topology in global models should be investigated.  Our choice to
impose a vertical field is motivated by the typical ``hourglass''
morphology \citep[e.g. figures~2 and 3 in][and references
  therein]{2010MNRAS.408..322K} emerging also in zoom-in
simulations of PPD formation \citep{2014IAUS..299..131N}, and observed
in molecular cores \citep{2006Sci...313..812G}; even though the
topologies of PPDs' magnetic fields have not yet been unambiguously
determined \citep{2014Natur.514..597S,0004-637X-797-2-74}.

% ------------------------------------------------------------------------------

\section{Conclusions}
\label{sec:concl}

The results presented in this paper, combined with those by \BS,
indicate that the traditional picture of a PPD hosting a dead zone in
the shielded disk interior, with turbulent surface layers driven by
the MRI, no longer holds. When ambipolar diffusion is treated, the
flow becomes laminar, and accretion is driven by the magnetic stresses
produced in launching a wind from the disk surface. Adding the Hall
effect appears to further modify the flow by allowing a significant
Maxwell stress to develop near the disk midplane through the combined
action of field winding and Hall drift \citep{2008MNRAS.385.1494K,%
  2013MNRAS.434.2295K,2014A&A...566A..56L,2014ApJ...791..137B}, but
does not lead to turbulence.  We are then left with a picture of PPDs
in which the innermost regions sustain vigorous MRI-induced turbulence
because of collisional ionization of alkali metals, the intermediate
regions between $\sim 0.5-20\au$ are laminar with angular momentum
transport occurring through a disk wind and perhaps through Maxwell
stresses near the midplane (an effect that only arises if ${\bf
  \Omega} \cdot {\bf B}>0$ and when including the Hall term), and an
outer region where the Hall effect and Ohmic resistivity are
sub-dominant and weak turbulence arises
\citep{2013ApJ...775...73S,2015ApJ...798...84B}.

The picture painted above has important implications for dust
evolution and planet formation. In principle, dust coagulation and
settling in the laminar region should occur rapidly, and in the
absence of a source of stirring the loss of small grains should affect
the spectral energy distribution (SED).  This was examined by
\citet{2005A&A...434..971D}, who showed that this rapid evolution of
the SED is not in agreement with observations, and for many years this
important result has provided indirect evidence for the existence of
turbulent stirring in the planet forming regions of protoplanetary
disks. In the absence of a magnetic origin for local turbulence this
suggests that either another source of turbulence exists, such as the
vertical shear instability \citep*[e.g.][]{2013MNRAS.435.2610N}, or
that small grains are continuously delivered to the laminar region
from the outer and/or inner turbulent regions.

The results in this paper are consistent  with the remanent magnetic
field measurements in chondrules by \citet{fu2014} in that the minimum 
midplane field strength for a wind driven disk at the presumed chondrule 
forming location (the midplane at $2$~to~$3 \au$) we find is smaller 
than the upper bound on the ambient field strength in the chondrule 
cooling region found in that work. This allows the chondrules studied 
by \citet{fu2014} to have formed in the midplane of a wind driven 
laminar disk, while also allowing for the chondrule formation process 
to locally concentrate the magnetic field.

The influence of the traditional dead-zone picture for protoplanetary
disks during various stages in planet formation has been examined in
some detail. For example,
\citet*{2011MNRAS.415.3291G,2012MNRAS.422.1140G} showed that the
stirring of planetesimals by the propagation of waves from the active
region into the dead zone can stir up small planetesimals ($\lesssim
10 \km$) and cause them to undergo collisional destruction, or prevent
their runaway growth. In recent work (Nelson et al. 2014, in prep.),
it has furthermore been shown that low mass planets orbiting in disk
models with traditional dead zones are likely to undergo very rapid
inward migration because the weak stresses there are unable to prevent
saturation of the corotation torque. In principle, the new emerging
picture of PPDs -- that is, assuming that Maxwell stresses operate
near the midplane because of the Hall effect -- can ameliorate these
problems by removing the turbulent stirring, and by providing a
significant stress that can prevent corotation torque saturation. Full
simulations are clearly required to examine whether or not the latter
effect can be realized during full nonlinear evolution. Finally,
\citet{2013ApJ...779...59G} examined gap formation and gas accretion
onto a giant planet within a disk with a traditional dead zone, and
observed that the gas in the gap region became enlivened because of
enhanced penetration of X-rays and cosmic rays. This created a
turbulent and chaotic gas flow into the planet Hill sphere, and the
enhanced magnetic coupling also led to the launching of a
magnetocentrifugally accelerated jet from the circumplanetary disk.
It remains unclear whether or not these effects will arise when
ambipolar diffusion is included, given the low density of the gap
region.

% ------------------------------------------------------------------------------

\acknowledgments 

We thank Martin Pessah, Anna Tenerani and Marco Velli for useful
advice on plasma instabilities, and the referee for a helpful report.
The research was carried out in part at the Jet Propulsion Laboratory,
California Institute of Technology, under a contract with the National
Aeronautics and Space Administration and with the support of the NASA
Origins of Solar Systems program via grant 13-OSS13-0114. The research
leading to these results has received funding from the Danish Council
for Independent Research (DFF) and FP7 Marie Curie Actions -- COFUND
under the grant-ID: DFF -- 1325-00111, and from the People Programme
(Marie Curie Actions) of the European Union's Seventh Framework
Programme (FP7/2007-2013) under REA grant agreement 327995.  This work
used the \NIII code developed by Udo Ziegler at the Leibniz Institute
for Astrophysics (AIP).  We acknowledge that the results of this
research have been partly achieved using the PRACE-3IP project (FP7
RI-312763) resource \texttt{Fionn} based in Ireland at the Irish
Centre for High-End Computing (ICHEC). Computations were also
performed on the \texttt{astro2} node at the Danish Center for
Supercomputing (DCSC), and on the U.K. STFC-funded \texttt{DiRAC}
facility.

% ------------------------------------------------------------------------------

\end{document}